\documentclass[fleqn,usenatbib]{mnras}

\usepackage{newtxtext,newtxmath}

\usepackage[T1]{fontenc}

\DeclareRobustCommand{\VAN}[3]{#2}
\let\VANthebibliography\thebibliography
\def\thebibliography{\DeclareRobustCommand{\VAN}[3]{##3}\VANthebibliography}

\usepackage{graphicx}	
\usepackage{amsmath}	

\allowdisplaybreaks

\title[Pop. effect on NIR mag and colour of RC stars]{The age and metallicity dependence of the near-infrared absolute magnitude and colour of red clump stars}

\author[H. Onozato et al.]{
Hiroki Onozato,$^{1}$\thanks{E-mail: hiroki.onozato@nao.ac.jp (HO)}
Yoshifusa Ita,$^{2}$
Yoshikazu Nakada$^{3}$
\\
$^{1}$National Astronomical Observatory of Japan, 2-21-1, Osawa, Mitaka, Tokyo, 181-8588, Japan\\
$^{2}$Astronomical Institute, Graduate School of Science, Tohoku University, 6-3, Aramaki Aoba, Aoba-ku, Sendai, Miyagi 980-8578, Japan\\
$^{3}$Kiso Observatory, Institute of Astronomy, School of Science, The University of Tokyo, 10762-30 Mitake, Kiso-machi, Kiso-gun, Nagano 397-0101, Japan
}

\date{Accepted XXX. Received YYY; in original form ZZZ}

\pubyear{\the\year{}}

\begin{document}
\label{firstpage}
\pagerange{\pageref{firstpage}--\pageref{lastpage}}
\maketitle

\begin{abstract}
Understanding the age and metallicity dependence of the absolute magnitude and colour of red clump (RC) stars is crucial for validating the accuracy of stellar evolution models and enhancing their reliability as a standard candle. However, this dependence has previously been investigated in the near-infrared across multiple bands only for -1.05 $\leq$ [Fe/H] $\leq$ 0.40, a range accessible through the star clusters in the Large Magellanic Cloud. Therefore, we used star clusters in the Small Magellanic Cloud and the Milky Way Galaxy to investigate the age and metallicity dependence of the absolute magnitude and colour of RC stars in the near-infrared for a broader parameter space (0.45 $\leq$ Age (Gyr) $\leq$ 10.5, -1.65 $\leq$ [Fe/H] $\leq$ 0.32). Comparison of our results with three isochronous models BaSTI, PARSEC, and MIST reveals that the age dependence of the absolute magnitude for young RC stars aligns well with theoretical predictions, within the fitting errors of the multiple regression analysis. Additionally, the observed colour shows good agreement with the theoretical models. Notably, the $J - K_{S}$ colour, which spans a wide parameter space, reproduces the distribution expected from the theoretical model.
\end{abstract}

\begin{keywords}
stars: distances -- Hertzsprung–Russell and colour–magnitude diagrams -- globular clusters: general -- open clusters and associations: general -- Magellanic Clouds
\end{keywords}



\section{Introduction}

Red clump (RC) stars are core helium-burning stars that have evolved from relatively metal-rich low-mass stars. These stars are widely used as a standard candle for determining the structure of the Milky Way Galaxy and measuring the distance to nearby galaxies. Additionally, RC stars serve as a "standard crayon" for investigating interstellar extinction due to their small variations in luminosity and colour, as well as their significant population density.

However, it is now established that the absolute magnitude and colour of RC stars exhibit an age and metallicity dependence (population effect), albeit not a strong one. Theoretically, the population effect of RC stars has been investigated by \citet{GS2001} and \citet{SG2002}. A precise understanding of this population effect is crucial for validating the accuracy of stellar evolution models and for employing it as a more reliable standard candle with population effect corrections. Observational verification of the population effect remains limited because of the difficulty in determining the age of RC stars.

One approach to estimating the age of RC stars is to use star clusters. The age of a star cluster can be estimated by comparing its colour-magnitude diagram with isochrones derived from theoretical models. Several studies have used star clusters in the Milky Way Galaxy to investigate the population effects on the absolute magnitude of RC stars \citep{GS2002, PS2003, vHG2007}. \citet{OIN2019} used star clusters in the Large Magellanic Cloud (LMC) to study the population effects of absolute magnitude and colour of RC stars. Another method for estimating the age of RC stars is using asteroseismology, although this approach is associated with significant age uncertainties \citep{CCZ2017}.

In this study, we analyse the colour of RC stars using star clusters from the Small Magellanic Cloud (SMC) and the Magellanic Bridge (MB), as well as the absolute magnitude and colour of RC stars using star clusters in the Milky Way Galaxy in addition to the LMC results of \citet{OIN2019}. The target clusters differ from those in the LMC in terms of age and metallicity: the clusters in the SMC and MB are older and have lower metallicity, while the open clusters in the Milky Way Galaxy have higher metallicity, thereby expanding the parameter space for investigating population effects. Additionally, the star clusters in each galaxy or region have an age-metallicity relation, where younger clusters have higher metallicity and older clusters have lower metallicity, making it difficult to study the effects of age and metallicity independently. However, because the SMC and MB have extended structures along the line of sight and cannot be considered equidistant, unlike the star clusters in the LMC, only the colour of the star clusters in the SMC is analysed.

Section~\ref{sec:data} describes the data and Section~\ref{sec:method} presents methods employed to select the star clusters and to determine the absolute magnitude and colour of the RC stars. Section~\ref{sec:res_and_dis} discusses the results obtained from these methods, while Section~\ref{sec:conclusions} provides the conclusions.

\section{The Data}\label{sec:data}
\subsection{SMC star clusters data}
We used the catalogue of \citet{BWK2020} as a list of star clusters in the SMC and the MB. This catalogue includes star clusters, associations, and related extended objects in the SMC and the MB. We chose objects classified as type C, denoting resolved star clusters, as the sample for this study. Among these, 117 star clusters have both age and metallicity measurements.

The near-infrared (NIR) magnitude of the individual members of the star clusters was obtained from the VISTA survey of the Magellanic Clouds (VMC survey) data release (DR) 5.1 \citep{CCG2011}. The VMC survey is a NIR survey in the $Y$, $J$, and $K_{s}$ bands of the Magellanic Cloud system conducted with the Visible and Infrared Survey Telescope for Astronomy \citep[VISTA telescope;][]{EMS2006} and equipped with the VISTA infrared camera \citep[VIRCAM;][]{DCW2006}. Data for the SMC, the MB and the Magellanic Stream are available in the VMC DR5.1. Given the crowded nature of these regions, we used the catalogue of point spread function (PSF) fitting photometry. Of the 117 star clusters listed in the catalogue of \citet{BWK2020}, 109 are included in the VMC survey area.

\subsection{Milky Way star clusters data}
The star clusters catalogue of \citet{vHG2007} was employed to select the target star clusters in the Milky Way Galaxy. Near-infrared photometric data were acquired from the Two Micron All Sky Survey \citep[2MASS, ][]{SCS2006}. The 2MASS is a near-infrared $J$, $H$, and $K_{S}$-band survey conducted using 1.3-m telescopes at Mount Hopkins, Arizona and Cerro Tololo, Chile. The 10~$\sigma$ detection limits are 15.8, 15.1, 14.3~mag at the $J$, $H$, $K_{S}$ bands, respectively, which are adequate for Milky Way star clusters. Age, metallicity, and $E(B-V)$ of the star clusters were taken from \citet{KPS2013}. The \textit{true} distance moduli (DM) of the star clusters were determined from the mean parallax of the member stars ($\omega$, the unit is arcsec), as provided by in \citet{CA2020} using the formula:
\begin{equation}
    DM = 5\log\omega - 5.
\end{equation}

\section{Method}\label{sec:method}
\subsection{Determination of the absolute magnitude and colour of the RC stars in each cluster}
\subsubsection{SMC star clusters}
\begin{figure*}
    \includegraphics{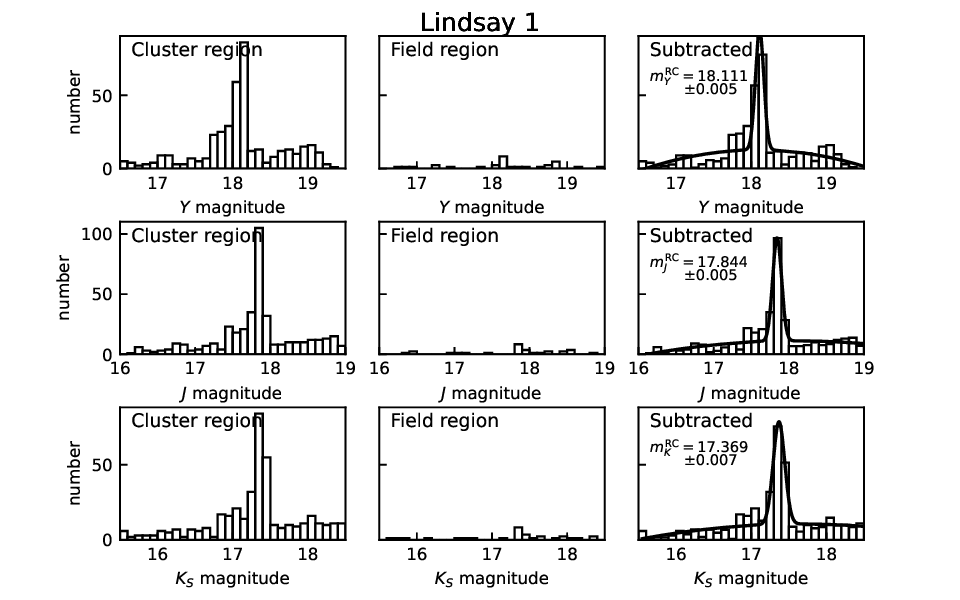}
    \includegraphics{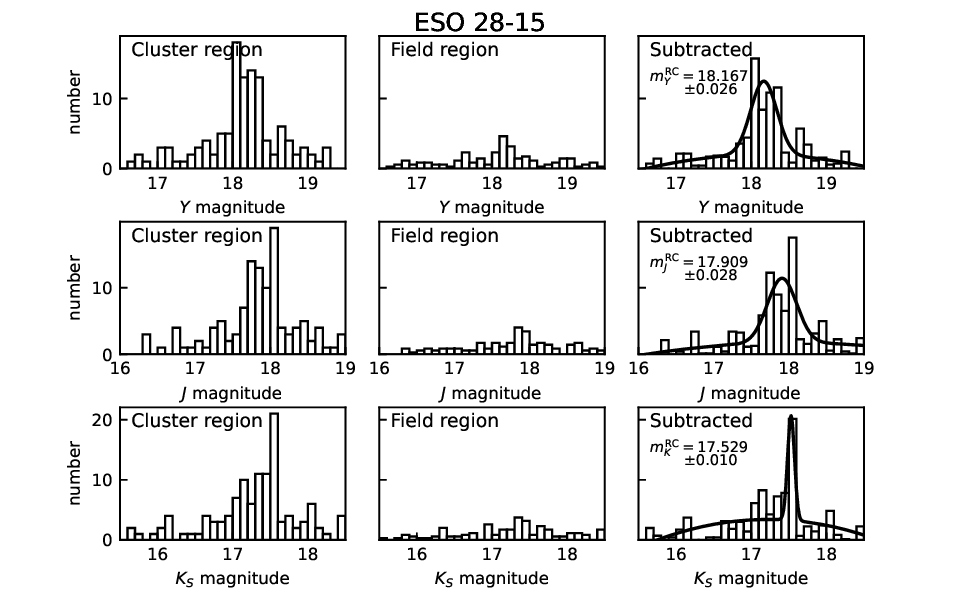}
    \caption{The luminosity functions of the RC stars as a function of their magnitude in the $Y$- (top), $J-$ (centre), and $K_{S}$-bands (bottom) for Lindsay\,1 and ESO\,28-15. (left) The luminosity functions of the cluster region. (centre) The luminosity functions of the field region. (right) The luminosity functions of the star cluster. The luminosity functions of the field region are subtracted from that of the cluster region. The fitting results with the equation (\ref{eq:fitting}) are shown by red lines.}
    \label{fig:Hist_1}
\end{figure*}

The colour of the RC stars was determined using a method similar to \citet{OIN2019}. First, stars in star clusters were selected based on the mean values of the semi-major and semi-minor axes described in \citet{BWK2020}. Figs.~\ref{fig:CMD} and \ref{fig:CMD_2} show the colour-magnitude diagrams (CMDs) for the selected stars of target star clusters in the SMC. We chose stars with $18.5 < m_{K_{S}} < 15.5$ and $0.2 < J - K_{S} < 0.8$ to determine the average colour of the RC stars in the star clusters. Next, we obtained the $m_{K_{S}}$ histograms for the rectangle region in CMDs containing RC stars as shown in Fig.~\ref{fig:Hist_1}. The control fields were defined as circular rings with inner radii equal to those of the star clusters and outer radii extending to the radii of the star cluster + 1.5\,arcmin. The $m_{K_{S}}$ histograms of the field stars were normalised to the ones that correspond to the same area as the target star clusters. Subsequently, these field $m_{K_{S}}$ histograms were subtracted from the cluster histograms, isolating the genuine cluster $m_{K_{S}}$ diagrams for analysis. The number of stars was normalised by the area of the star cluster when the histograms were created. For Lindsay\,1, the outer radius of the surrounding region was set to the star cluster radius + 1.2\,arcmin because of the discontinuity in the stellar density distribution at the border of the observed region. The apparent magnitude of selected stars were fitted with the following functional form as \citet{G2016}
\begin{equation}
N(m_{\lambda}) = a + bm_{\lambda}+cm_{\lambda}^2 + d \exp\left[-\frac{(m_{\lambda}^{RC} - m_{\lambda})^2}{2\sigma_{\lambda}^2}\right],
\label{eq:fitting}
\end{equation}
where $\lambda$ is a passband ($YJK_{S}$). The quadratic term represents the distribution of red giant branch stars, while the Gaussian term corresponds to the distribution of RC stars. $m_{\lambda}^{\mathrm{RC}}$ is the mean apparent magnitude and $\sigma_{\lambda}$ is the standard deviation of the RC stars. The number of RC stars ($N_{\mathrm{RC}}$) to calculate standard errors are given by
\begin{equation}
N_{\mathrm{RC}} = \sqrt{2\pi} \sigma_{\lambda} \times 10d.
\end{equation}
A total of 28 star clusters could be fitted using this formula. The number of star clusters in each selection process is shown in Table~\ref{tab:number}. We corrected interstellar extinction using $E(B - V)$ values of \citet{BWK2020}, as shown in Table~\ref{tab:cluster}, and the extinction law of \citet{CCM1989}. We adopted the $R_{V}$ value of 3.1. The colour of the RC stars for each cluster was derived by subtracting the magnitude in each passband determined through this procedure.

\begin{figure*}
    \includegraphics{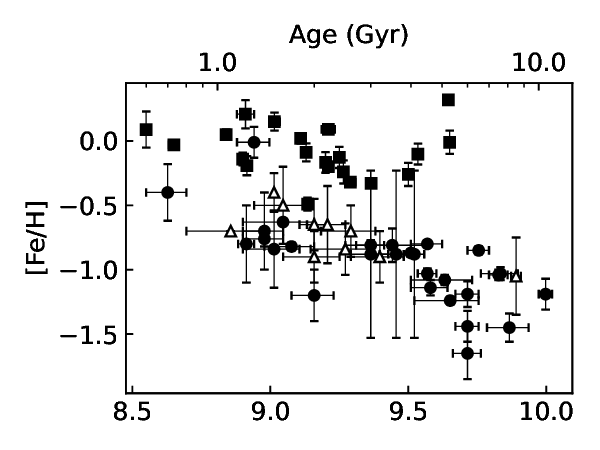}
    \caption{The age and metallicity of star clusters used to derive the absolute magnitude or colour of RC stars. Black circles are our SMC samples, black squares are our MW samples, and grey triangles are the samples of \citet{OIN2019}.}
    \label{fig:Age_and_Metallicity}
\end{figure*}

Information on the star clusters for which the colour of RC stars has been derived is summarised in Table~\ref{tab:cluster}. Fig.~\ref{fig:Age_and_Metallicity} shows the age and metallicity distribution of the star clusters for which the absolute magnitude or colour of the RC stars has been determined in this work and previous studies. It can be seen that our sample includes many metal-poor and old star clusters that have not been previously covered.


\begin{figure*}
    \includegraphics{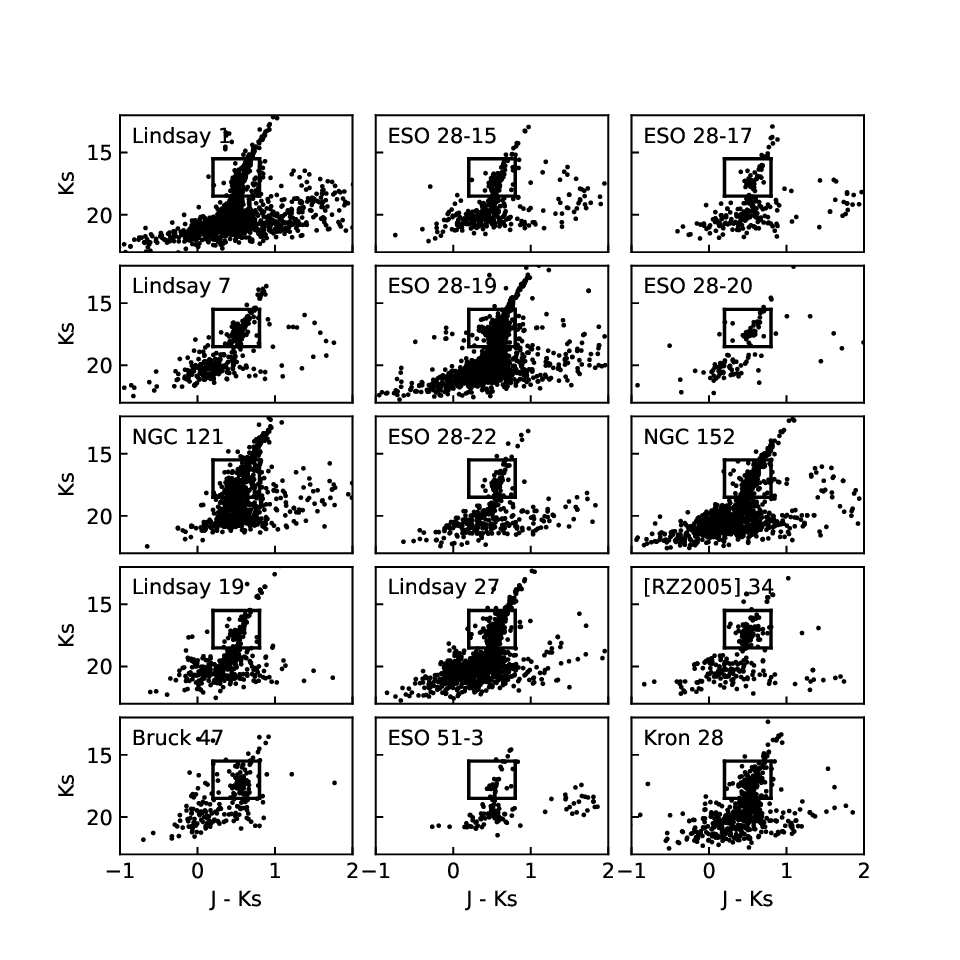}
    \caption{$J - K_{S}$ versus $K_{S}$ colour magnitude diagrams of our target star clusters of the SMC. The black rectangles represent the boundary to select the stars for fitting.}
    \label{fig:CMD}
\end{figure*}

\begin{figure*}
    \includegraphics{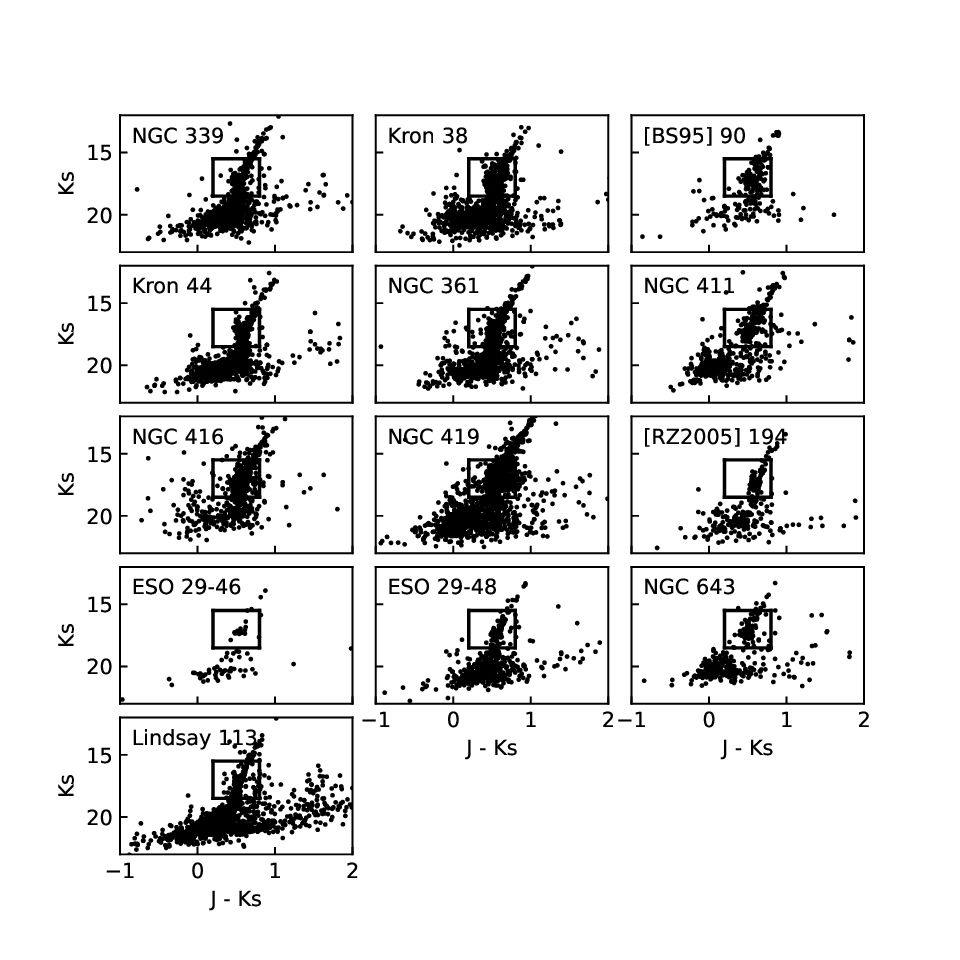}
    \caption{Continued from Fig.~\ref{fig:CMD}.}
    \label{fig:CMD_2}
\end{figure*}

\begin{table*}
\caption{The number of star clusters at each selection process}
\label{tab:number}
\begin{tabular}{lc}
\hline
Selection process & number of clusters\\
\hline
All objects in \citet{BWK2020} & 2741\\
All star clusters (Type of object is C) in \citet{BWK2020} & 626\\
Star clusters with age and metallicity & 117\\
Star clusters in the VMC survey region & 109\\
Star clusters that have significant RC excess and can be fitted by equation~(\ref{eq:fitting}) & 28\\
\hline
\end{tabular}
\end{table*}

\begin{table*}
\caption{SMC Cluster Information}
\label{tab:cluster}
\scalebox{0.9}{
\begin{tabular}{lcccccccccc}
\hline
Cluster name & RA (J2000.0) & Dec (J2000.0) & Radius (arcmin) & Age (Gyr) & [Fe/H] & $E(B - V)$ & Age Ref.$^a$ & [Fe/H] Ref.$^a$ & $E(B - V)$ Ref.$^a$\\
\hline
Lindsay\,1 & 00$^{\mathrm{h}}$ 03$^{\mathrm{m}}$ 54\fs6 & -73\degr 28\arcmin 16\arcsec & 2.3  & $7.5 \pm 0.5$  & $-1.04 \pm 0.04$ & 0.019$^b$ & (1) & (2) & (1) \\
ESO\,28-15 & 00$^{\mathrm{h}}$ 21$^{\mathrm{m}}$ 27\fs3 & -73\degr 44\arcmin 53\arcsec & 1.1 & $5.1^{+2.8}_{-1.8}$ $ ^{c}$ & $-1.08 \pm 0.04$ & 0.04 & (3)(4)$^{d}$ & (5) & (6)\\
ESO\,28-17 & 00$^{\mathrm{h}}$ 23$^{\mathrm{m}}$ 04\fs0 & -73\degr 40\arcmin 12\arcsec & 0.85 & $5.3^{+3.4}_{-2.0}$ $^{c}$ & $-1.24 \pm 0.03$ & 0.03 & (3)(6) & (5) & (6)\\
Lindsay\,7 & 00$^{\mathrm{h}}$ 24$^{\mathrm{m}}$ 43\fs16 & -73\degr 45\arcmin 11\farcs7 & 0.9 & 1.4$^{e}$ & $-0.76 \pm 0.06$ & 0.02 & (3)(6) & (5) & (6)\\
ESO\,28-19 & 00$^{\mathrm{h}}$ 24$^{\mathrm{m}}$ 46\fs0 & -72\degr 47\arcmin 38\arcsec & 1.7 & $6.5 \pm 0.5$ & $-0.85 \pm 0.03 $ & 0.019$^b$ & (1) & (2) & (1) \\
ESO\,28-20 & 00$^{\mathrm{h}}$ 25$^{\mathrm{m}}$ 26\fs60 & -74\degr 04\arcmin 29\farcs7 & 0.50 & $1.6 \pm 0.4$ & $-0.63 \pm 0.02 $ & 0.03 & (7) & (2) & (7) \\
NGC\,121 & 00$^{\mathrm{h}}$ 26$^{\mathrm{m}}$ 48\fs5 & -71\degr 32\arcmin 05\arcsec & 1.725 & $10.5 \pm 0.5$ & $-1.19 \pm 0.12 $ & 0.027$^b$ & (8) & (9) & (8) \\
ESO\,28-22 & 00$^{\mathrm{h}}$ 27$^{\mathrm{m}}$ 45\fs17 & -72\degr 46\arcmin 52\farcs5 & 0.85 & $3.5 \pm 1.0$ & $-0.81 \pm 0.13 $ & 0.040 & (10) & (9) & (10) \\
NGC\,152 & 00$^{\mathrm{h}}$ 32$^{\mathrm{m}}$ 56\fs3 & -73\degr 06\arcmin 57\arcsec & 1.5 & $1.23 \pm 0.07$ & $-0.80 \pm 0.30 $ & $0.03 \pm 0.01$ & (4) & (9) & (4) \\
Lindsay\,19 & 00$^{\mathrm{h}}$ 37$^{\mathrm{m}}$ 41\fs78 & -73\degr 54\arcmin 19\farcs7 & 0.85 & 4.0$^e$ & $-0.87 \pm 0.03$ & 0.02 & (3)(6) & (5) & (6)\\
Lindsay\,27 & 00$^{\mathrm{h}}$ 41$^{\mathrm{m}}$ 24\fs2 & -72\degr 53\arcmin 27\arcsec & 1.25 & $4.6 \pm 0.6$ & $-1.14 \pm 0.06$ & 0.11 & (3) & (5) & (6)\\
\lbrack RZ2005\rbrack\,34 & 00$^{\mathrm{h}}$ 42$^{\mathrm{m}}$ 59\fs87 & -72\degr 35\arcmin 21\farcs0 & 0.50 & $0.70 \pm 0.10$ & $-0.40 \pm 0.22$ & $0.000 \pm 0.010$ & (3) & (11) & (11)\\
Bruck\,47 & 00$^{\mathrm{h}}$ 48$^{\mathrm{m}}$ 33\fs23 & -73\degr 18\arcmin 25\farcs1 & 0.50 & $1.29_{-1.9}^{+1.1}$ $^c$ & $-0.01 \pm 0.12$ & $0.080 \pm 0.030$ & (3)(12) & (11) & (11)\\
ESO\,51-3 & 00$^{\mathrm{h}}$ 48$^{\mathrm{m}}$ 50\fs0 & -69\degr 52\arcmin 12\arcsec & 0.90 & $6.0_{-0.4}^{+0.5}$ $^{c}$ & $-1.65 \pm 0.20 $ & 0.013$^b$ & (1)(3)(13) & (13) & (1) \\
Kron\,28 & 00$^{\mathrm{h}}$ 51$^{\mathrm{m}}$ 39\fs55 & -71\degr 59\arcmin 56\farcs6 & 0.85 & $2.0 \pm 0.3$ & $-1.20 \pm 0.20$ & 0.06 & (3) & (13) & (13) \\
NGC\,339 & 00$^{\mathrm{h}}$ 57$^{\mathrm{m}}$ 47\fs5 & -74\degr 28\arcmin 17\arcsec & 1.45 & $6.0 \pm 0.5$ & $-1.19 \pm 0.10 $ & 0.032 & (1) & (9) & (1) \\
Kron\,38 & 00$^{\mathrm{h}}$ 57$^{\mathrm{m}}$ 49\fs5 & -73\degr 25\arcmin 23\arcsec & 1.1 & $3.0 \pm 0.4$ & $-0.88 \pm 0.65$ & $0.010 \pm 0.020$ & (3) & (11) & (11) \\
\lbrack BS95\rbrack\,90 & 00$^{\mathrm{h}}$ 59$^{\mathrm{m}}$ 07\fs25 & -72\degr 08\arcmin 59\farcs1 & 0.50 & $4.5 \pm 0.5$ & $-0.80^{f}$ & 0.021 & (14) & (15) & (14)\\
Kron\,44 & 01$^{\mathrm{h}}$ 02$^{\mathrm{m}}$ 04\fs0 & -73\degr 55\arcmin 33\arcsec & 1.45 & $3.0 \pm 0.3$ & $-0.81 \pm 0.04$ & 0.05 & (3) & (2) & (13) \\
NGC\,361 & 01$^{\mathrm{h}}$ 02$^{\mathrm{m}}$ 11\fs0 & -71\degr 36\arcmin 21\arcsec & 1.3 & $8.10 \pm 1.20$ & $-1.45 \pm 0.11 $ & $0.07 \pm 0.03$ & (16) & (16) & (16) \\
NGC\,411 & 01$^{\mathrm{h}}$ 07$^{\mathrm{m}}$ 55\fs3 & -71\degr 46\arcmin 04\arcsec & 1.05 & $1.5 \pm 0.3$ $^{g}$ & $-0.84 \pm 0.30 $ & 0.03 & (7)(17)(18)(19) & (9) & (7) \\
NGC\,416 & 01$^{\mathrm{h}}$ 07$^{\mathrm{m}}$ 59\fs0 & -72\degr 21\arcmin 20\arcsec & 0.85 & $6.0 \pm 0.5$ & $-1.44 \pm 0.12 $ & $0.08 \pm 0.03$ & (1) & (16) & (16) \\
NGC\,419 & 01$^{\mathrm{h}}$ 08$^{\mathrm{m}}$ 18\fs0 & -72\degr 53\arcmin 02\arcsec & 1.4 & $1.4 \pm 0.2$ & $-0.70 \pm 0.30 $ & 0.083$^b$ & (1) & (9) & (1) \\
\lbrack RZ2005\rbrack\,194 & 01$^{\mathrm{h}}$ 12$^{\mathrm{m}}$ 51\fs74 & -73\degr 07\arcmin 10\farcs9 & 0.60 & $4.1 \pm 0.3$ & $-0.88 \pm 0.65$ & $0.040 \pm 0.010$ & (3) & (11) & (20)\\
ESO\,29-46 & 01$^{\mathrm{h}}$ 33$^{\mathrm{m}}$ 14\fs0 & -74\degr 10\arcmin 00\arcsec & 0.60 & $3.6^{+0.4}_{-0.3}$ $^c$ & $-0.88 \pm 0.65 $ & $0.010 \pm 0.005$ & (3)(21) & (11) & (11) \\
ESO\,29-48 & 01$^{\mathrm{h}}$ 34$^{\mathrm{m}}$ 26\fs0 & -72\degr 52\arcmin 28\arcsec & 1.45 & $7.6 \pm 1.0$ & $-1.03 \pm 0.05 $ & 0.06 & (3) & (5) & (21) \\
NGC\,643 & 01$^{\mathrm{h}}$ 35$^{\mathrm{m}}$ 01\fs0 & -75\degr 33\arcmin 23\arcsec & 1.1 & $1.7 \pm 0.3^{b}$ & $-0.82 \pm 0.03 $ & 0.07 & (3)(21) & (5) & (22)(23) \\
Lindsay\,113 & 01$^{\mathrm{h}}$ 49$^{\mathrm{m}}$ 30fs3 & -73\degr 43\arcmin 40\arcsec & 2.2 & $4.4^{+0.6}_{-0.5}$ $^{b}$ & $-1.03 \pm 0.04$ & $0.030 \pm 0.010$ & (3)(21) & (2) & (24)\\
\hline
\multicolumn{10}{l}{$^a$ (1) \citet{GGS2008} (2) \citet{PGC2015} (3) \citet{PGC2014} (4) \citet{DKB2016} (5) \citet{PGG2009} (6) \citet{PSG2005} (7) \citet{PSC2005}}\\
\multicolumn{10}{l}{\quad (8) \citet{GGG2008} (9) \citet{DH1998} (10) \citet{MJD1992} (11) \citet{PPV2017} (12) \citet{P2011b} (13) \citet{PSC2001}} \\
\multicolumn{10}{l}{\quad (14) \citet{RGB2007} (15) \citet{SSN2007} (16) \citet{MSF1998} (17) \citet{AS1999} (18)  \citet{DM1986}}\\
\multicolumn{10}{l}{\quad (19) \citet{dFBI1998} (20) \citet{P2011a} (21) \citet{PdGR2015} (22) \citet{PCB2011} (23) \citet{PSG2007b} (24) \citet{PSG2007a}}\\
\multicolumn{10}{l}{$^b$ Calculated from $E(V - I)$ using \citet{CCM1989}'s law.}\\
\multicolumn{10}{l}{$^c$ Logarithmic mean of the age of the references.}\\
\multicolumn{10}{l}{$^d$ (5) refers to (7), but their ages are slightly different ((5) 3.3\>Gyr, (7) 3.1\>Gyr).}\\
\multicolumn{10}{l}{$^e$ The way \citet{BWK2020} derived the age is unclear (Lindsay\,7: (3) $1.6 \pm 0.2$\>Gyr, (6) 2.0\>Gyr; Lindsay\,19: (3) $4.8 \pm 0.7$\>Gyr, (6) 2.1\>Gyr).}\\
\multicolumn{10}{l}{$^f$ The metallicity is given in the form of $Z$ ($Z = 0.002$).}\\
\multicolumn{10}{l}{$^g$ Average of the age of the references.}\\
\end{tabular}
}
\end{table*}

\subsubsection{Milky Way star clusters}
The absolute magnitude and colour of RC stars in star clusters containing a sufficient number of RC stars were determined using a method similar to that used for the SMC star clusters. The differences are as follows. First, we selected stars with $(DM - 3.5) < m_{K_{S}} < (DM - 0.5)$ and $0.4 < J - K_{S} < 1.0$ to derive the absolute magnitude and colour of the RC stars as shown in Figs.~\ref{fig:MWSC_CMD} and \ref{fig:MWSC_CMD_2}. Second, the members of the star clusters were selected based on the membership probability of \citet{CA2020}. We considered stars with a membership probability greater than 0.9 to be members of the star clusters, and we did not subtract field stars. For star clusters with fewer stars, the absolute magnitude and colour of RC stars were determined by calculating the average for stars within the same magnitude and colour range. Thus, the absolute magnitude and colour could be determined for the clusters studied by \citet{vHG2007} excluding NGC\,2420 and NGC\,6633, both of which contain only two RC stars. Only colour was used for Haffner\,2, as its large DM of 15.512\,mag (12.658\,kpc) reduces the reliability of Gaia parallax measurements. A list of star clusters of this study is presented in Table~\ref{tab:MW_cluster}.

\begin{table*}
\caption{Milky Way Cluster Information}
\label{tab:MW_cluster}
\begin{tabular}{lcccccc}
\hline
Cluster name & RA (J2000.0) & Dec (J2000.0) & Age ($\log t$) & [Fe/H] & $E(B - V)$ & $DM$ \\
\hline
NGC\,188 & 00$^{\mathrm{h}}$ 47$^{\mathrm{m}}$ 11\fs52 & +85\degr 14\arcmin 38\farcs4 & 9.650 & $-0.010 \pm 0.090$ & 0.085 & $11.475 \pm 0.009$\\
NGC\,752 & 01$^{\mathrm{h}}$ 56$^{\mathrm{m}}$ 53\fs51 & +37\degr 47\arcmin 38\farcs6 & 9.130 & $-0.088 \pm 0.070$ & 0.040 & $8.250 \pm 0.005$\\
NGC\,1817 & 05$^{\mathrm{h}}$ 12$^{\mathrm{m}}$ 33\fs4 & +16\degr 41\arcmin 46\arcsec & 8.900 & $-0.140 \pm 0.050$ & 0.354 & $11.294 \pm 0.012$\\
M\,37 (NGC\,2099) & 05$^{\mathrm{h}}$ 52$^{\mathrm{m}}$ 17\fs8 & +32\degr 32\arcmin 42\arcsec & 8.550 & $0.089 \pm 0.140$ & 0.350 & $10.883 \pm 0.007$\\
NGC\,2204 & 06$^{\mathrm{h}}$ 15$^{\mathrm{m}}$ 31\fs7 & -18\degr 40\arcmin 12\arcsec & $9.290 \pm 0.018$ & $-0.320 \pm 0.100$ & 0.021 & $13.399 \pm 0.031$\\
NGC\,2243 & 06$^{\mathrm{h}}$ 29$^{\mathrm{m}}$ 34\fs8 & -31\degr 16\arcmin 55\arcsec & 9.135 & $-0.490 \pm 0.050$ & 0.062 & $13.379 \pm 0.031$\\
Haffner\,2 (Tombaugh\,2) & 07$^{\mathrm{h}}$ 03$^{\mathrm{m}}$ 05\fs5 & -20\degr 49\arcmin 12\arcsec & 9.010 & $-0.310 \pm 0.020$ & 0.354 & $15.512 \pm 0.165$\\
NGC\,2360 & 07$^{\mathrm{h}}$ 17$^{\mathrm{m}}$ 46\fs3 & -15\degr 37\arcmin 52\arcsec & $8.650 \pm 0.018$ & $-0.030 \pm 0.010$ & 0.416 & $10.884 \pm 0.005$\\
Melotte\,66 & 07$^{\mathrm{h}}$ 26$^{\mathrm{m}}$ 17\fs5 & -47\degr 41\arcmin 06\arcsec & 9.365 & $-0.330 \pm 0.030$ & 0.125 & $13.688 \pm 0.036$\\
Berkeley\,39 & 07$^{\mathrm{h}}$ 46$^{\mathrm{m}}$ 48\fs5 & -04\degr 39\arcmin 54\arcsec & 9.500 & $-0.260 \pm 0.090$ & 0.042 & $13.484 \pm 0.054$\\
NGC\,2477 & 07$^{\mathrm{h}}$ 52$^{\mathrm{m}}$ 11\fs0 & -38\degr 32\arcmin 13\arcsec & $8.915 \pm 0.007$ & $-0.192 \pm 0.074$ & 0.291 & $10.886 \pm 0.003$\\
NGC\,2506 & 08$^{\mathrm{h}}$ 00$^{\mathrm{m}}$ 02\fs4 & -10\degr 46\arcmin 23\arcsec & $9.210 \pm 0.017$ & $-0.200 \pm 0.020$ & 0.042 & $12.673 \pm 0.015$\\
NGC\,2527 & 08$^{\mathrm{h}}$ 04$^{\mathrm{m}}$ 59\fs0 & -28\degr 07\arcmin 19\farcs2 & $8.910 \pm 0.031$ & $-0.208 \pm 0.110$ & 0.040 & $9.068 \pm 0.006$\\
NGC\,2682 & 08$^{\mathrm{h}}$ 51$^{\mathrm{m}}$ 23\fs0 & +11\degr 48\arcmin 50\arcsec & $9.535 \pm 0.009$ & $-0.102 \pm 0.081$ & 0.050 & $9.725 \pm 0.004$\\
NGC\,3680 & 11$^{\mathrm{h}}$ 25$^{\mathrm{m}}$ 34\fs1 & -43\degr 14\arcmin 24\arcsec & 9.200 & $-0.167 \pm 0.080$ & 0.062 & $10.151 \pm 0.010$\\
NGC\,3960 & 11$^{\mathrm{h}}$ 50$^{\mathrm{m}}$ 34\fs6 & -55\degr 40\arcmin 44\arcsec & 9.110 & $0.020 \pm 0.040$ & 0.167 & $11.990 \pm 0.011$\\
NGC\,5852 & 15$^{\mathrm{h}}$ 04$^{\mathrm{m}}$ 12\fs2 & -54\degr 21\arcmin 58\arcsec & $8,840 \pm 0.013$ & $0.050 \pm 0.040$ & 0.312 & $9.628 \pm 0.004$\\
NGC\,6134 & 16$^{\mathrm{h}}$ 27$^{\mathrm{m}}$ 48\fs7 & -49\degr 09\arcmin 40\arcsec & $9.015 \pm 0.019$ & $0.150 \pm 0.070$ & 0.458 & $10.363 \pm 0.005$\\
IC\,4651 & 17$^{\mathrm{h}}$ 24$^{\mathrm{m}}$ 50\fs9 & -49\degr 55\arcmin 01\arcsec & 9.250 & $-0.128 \pm 0.082$ & 0.121 & $9.882 \pm 0.004$\\
NGC\,6791 & 19$^{\mathrm{h}}$ 20$^{\mathrm{m}}$ 53\fs0 & +37\degr 46\arcmin 41\arcsec & 9.645 & $0.320 \pm 0.020$ & 0.117 & $13.583 \pm 0.023$\\
NGC\,6819 & 19$^{\mathrm{h}}$ 41$^{\mathrm{m}}$ 18\fs5 & +40\degr 11\arcmin 24\arcsec & $9.210 \pm 0.024$ & $0.090 \pm 0.030$ & 0.237 & $12.243 \pm 0.006$\\
NGC\,7789 & 23$^{\mathrm{h}}$ 57$^{\mathrm{m}}$ 20\fs2 & +56\degr 43\arcmin 34\arcsec & $9.265 \pm 0.009$ & $-0.240 \pm 0.090$ & 0.237 & $11.720 \pm 0.005$\\
\hline
\end{tabular}
\end{table*}

\begin{figure*}
    \includegraphics{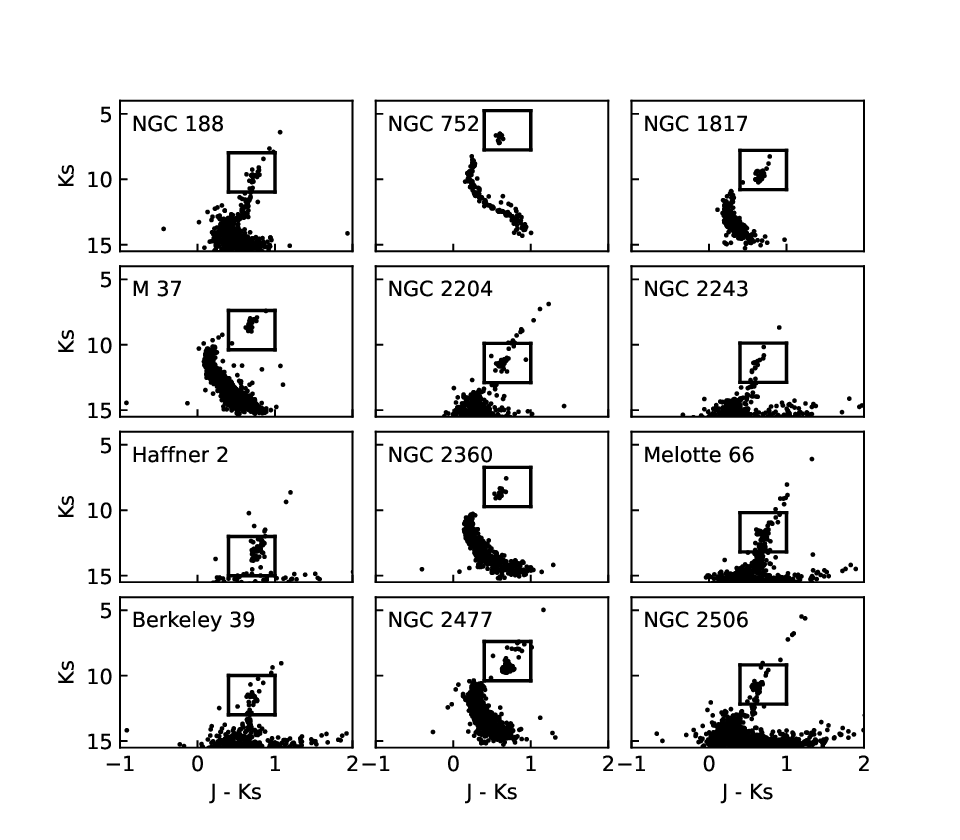}
    \caption{The same figure as Fig.~\ref{fig:CMD} but for Milky Way star clusters.}
    \label{fig:MWSC_CMD}
\end{figure*}

\begin{figure*}
    \includegraphics{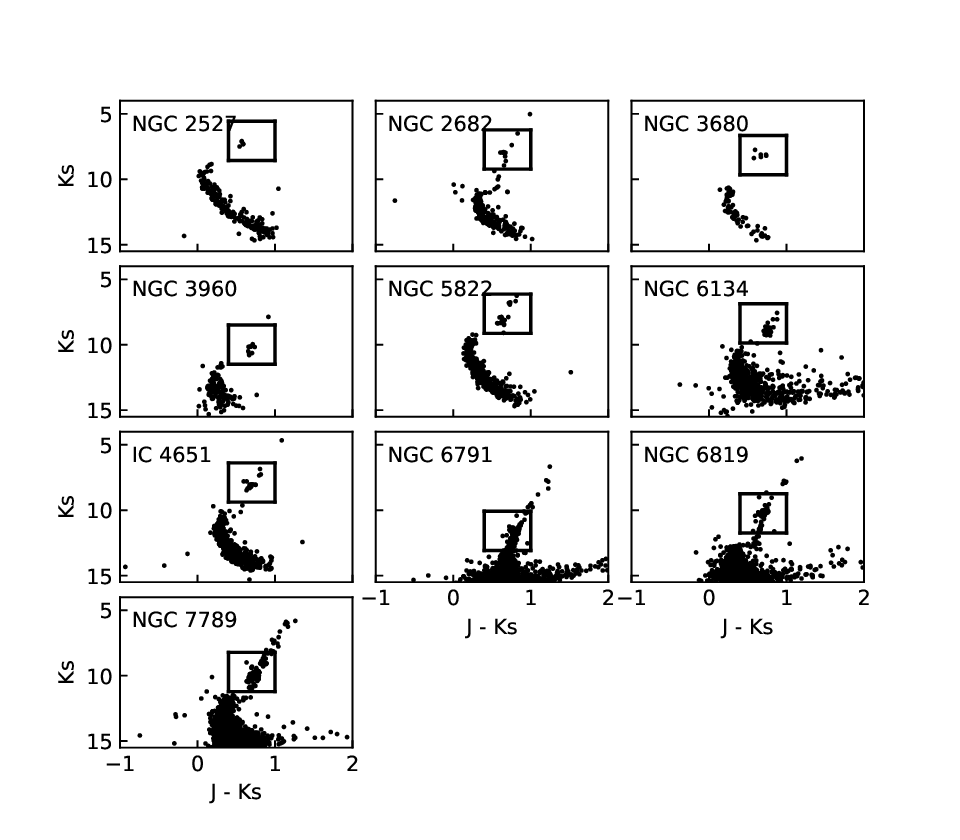}
    \caption{Continued from Fig.~\ref{fig:MWSC_CMD}.}
    \label{fig:MWSC_CMD_2}
\end{figure*}

\subsection{Multiple Regression Analysis}
We performed a multiple regression analysis, following the method of \citet{OIN2019} to confirm the population effects. This also enables correction for population effects when RC stars are used as a standard candle or crayon. The age range for the multiple regression analysis of absolute magnitude was restricted to 1--4 Gyr, as this range provided a sufficient sample size. The colour was analysed for the entire sample because the SMC dataset provided sufficient samples across the remaining range. To derive the correction formula, we performed least-squares fitting using the following function
\begin{equation}
M_{\lambda} = a (\log(t) - 9.300)  + b([\mathrm{Fe}/\mathrm{H}] + 0.50) + c,
\label{eq:AM_MRA}
\end{equation}
where $t$ represents the age (yr) of the star clusters. For colour, replace $M_{\lambda}$ with $m_{\lambda,\,1} - m_{\lambda,\,1}$.

\section{Results and Discussion}\label{sec:res_and_dis}
\begin{figure*}
    \includegraphics{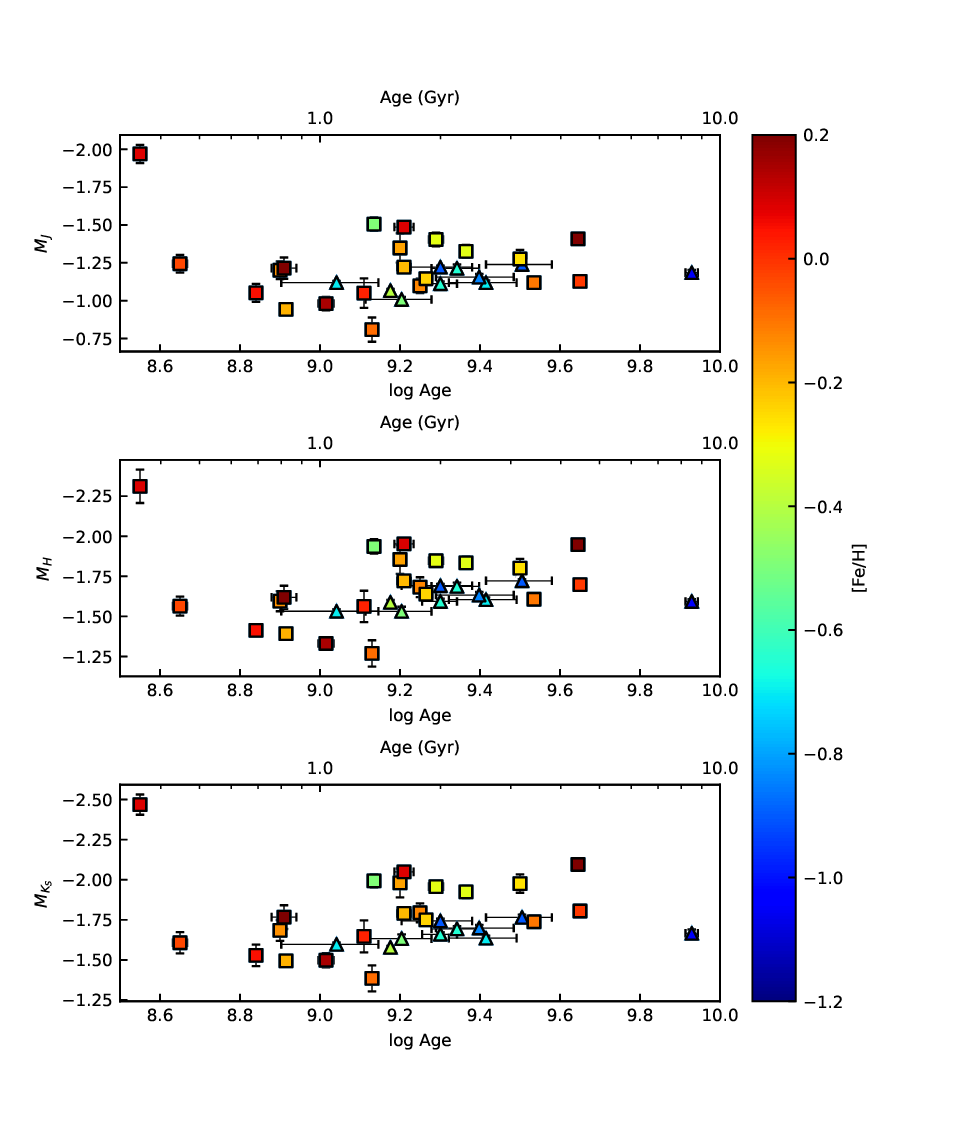}
    \caption{Mean RC absolute magnitude versus age in each cluster for $J$ (left), $H$ (centre), and $K_{S}$ band (right). Colour scale represents metallicity difference.}
    \label{fig:AM_age}
\end{figure*}

\begin{figure*}
    \includegraphics{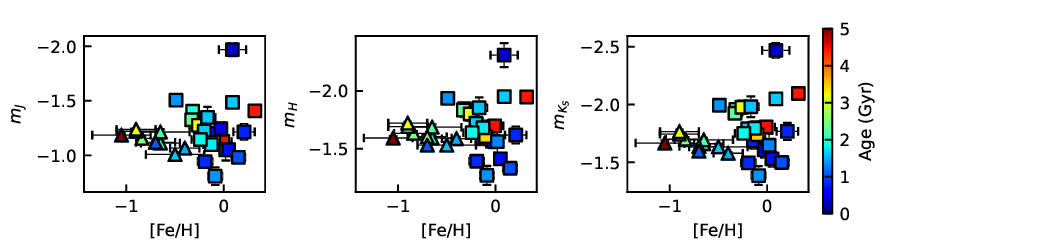}
    \caption{Mean RC absolute magnitude versus metallicity in each cluster for $J$ (left), $H$ (centre), and $K_{S}$ band (right). Colour scale represents age difference.}
    \label{fig:AM_metal}
\end{figure*}

\begin{figure*}
    \includegraphics{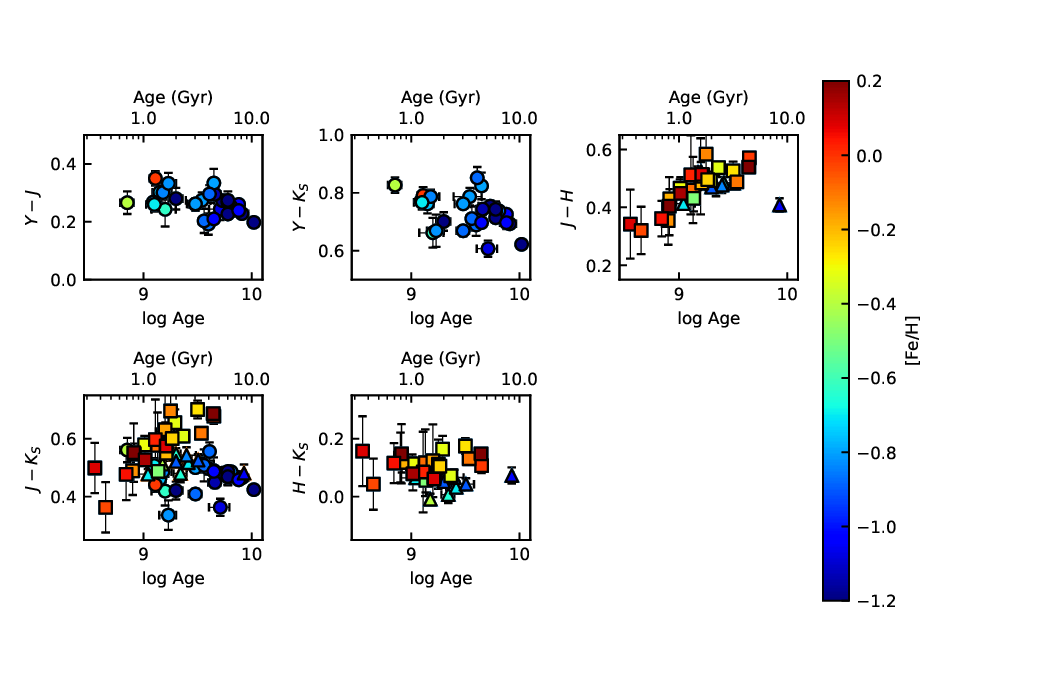}
    \caption{Mean RC colour versus age in each cluster for $Y - J$ (upper left), $Y - K_{S}$ (upper centre), $J - H$ band (upper right), $J - K_{S}$ (lower left), and $H - K_{S}$ (lower centre). Colour scale represents metallicity difference.}
    \label{fig:color_age}
\end{figure*}

\begin{figure*}
    \includegraphics{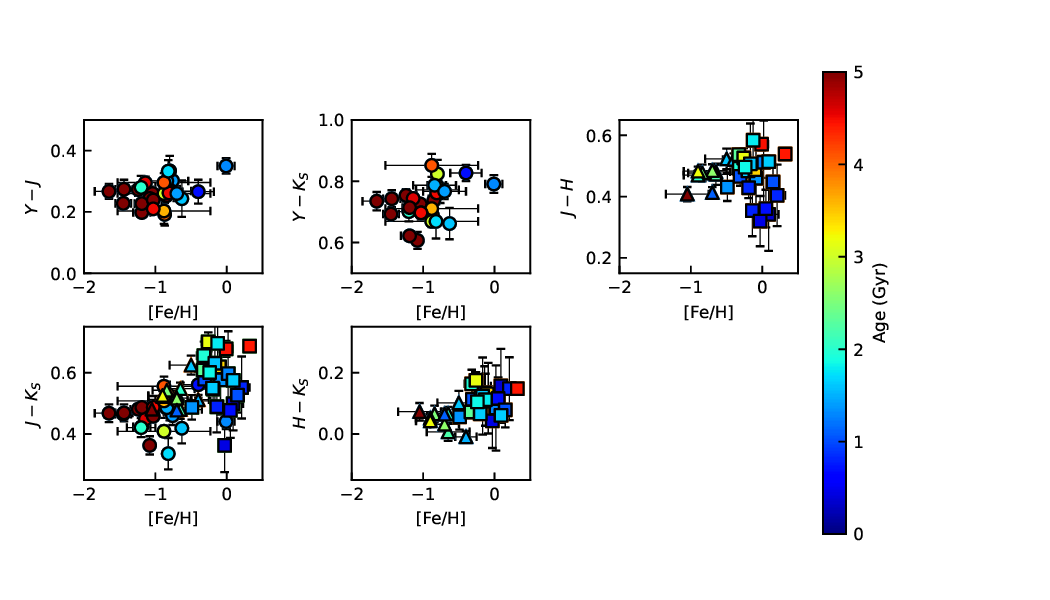}
    \caption{Mean RC colour versus metallicity in each cluster for $Y - J$ (upper left), $Y - K_{S}$ (upper centre), $J - H$ band (upper right), $J - K_{S}$ (lower left), and $H - K_{S}$ (lower centre). Colour scale represents age difference.}
    \label{fig:color_metal}
\end{figure*}

The derived absolute magnitude and intrinsic colour of RC stars in the star clusters are summarised in Tables~\ref{tab:RC_colors} and \ref{tab:RC_am_colors}. The absolute magnitude of the RC stars for each cluster, plotted against age and metallicity, are shown in Figs.~\ref{fig:AM_age} and \ref{fig:AM_metal}, respectively. Figs.~\ref{fig:color_age} and \ref{fig:color_metal} depict the plots for colour.

\begin{table}
\caption{The derived RC colour of the SMC target star clusters.}
\label{tab:RC_colors}
\begin{tabular}{lccc}
\hline
Cluster name & $Y - J$ & $Y - K_{S}$ & $J - K_{S}$\\
\hline
Lindsay\,1 & $0.259 \pm 0.007$ & $0.726 \pm 0.008$ & $0.466 \pm 0.008$\\
ESO\,28-15 & $0.244 \pm	0.038$ & $0.607 \pm 0.028$ & $0.362 \pm 0.030$\\
ESO\,28-17 & $0.273 \pm 0.021$ & $0.753 \pm 0.020$ & $0.481 \pm 0.018$\\
Lindsay\,7 & $0.302 \pm 0.036$ & $0.762 \pm 0.033$ & $0.460 \pm 0.031$\\
ESO\,28-19 & $0.251 \pm 0.013$ & $0.737 \pm	0.013$ & $0.486 \pm 0.012$\\
ESO\,28-20 & $0.243 \pm 0.060$ & $0.648 \pm 0.051$  & $0.405 \pm 0.050$\\
NGC\,121 & $0.199 \pm 0.013$ & $0.622 \pm 0.012$ & $0.423 \pm 0.012$\\
ESO\,28-22 & $0.274 \pm 0.028$ & $0.787 \pm 0.031$ & $0.513 \pm 0.030$\\
NGC\,152 & $0.268 \pm 0.015$ & $0.772 \pm 0.016$ & $0.504 \pm 0.015$\\
Lindsay\,19 & $0.193 \pm 0.036$ & $0.689 \pm 0.037$ & $0.496 \pm 0.021$\\
Lindsay\,27 & $0.295 \pm 0.019$ & $0.744 \pm 0.020$ & $0.449 \pm 0.019$\\
\lbrack RZ2005\rbrack\,34 & $0.266 \pm 0.039$ & $0.828 \pm 0.027$ & $0.562 \pm 0.042$\\
Bruck\,47 & $0.350 \pm 0.025$ & $0.791 \pm 0.029$ & $0.441 \pm 0.024$\\
ESO\,51-3 & $0.267 \pm 0.024$ & $0.735 \pm 0.030$ & $0.468 \pm 0.029$\\
Kron\,28 & $0.280 \pm 0.038$ & $0.702 \pm 0.032$ & $0.421 \pm 0.031$\\
NGC\,339 & $0.229 \pm 0.011$ & $0.714 \pm 0.012$ & $0.485 \pm 0.011$\\
Kron\,38 & $0.280 \pm 0.022$ & $0.720 \pm 0.019$ & $0.440 \pm 0.018$\\
\lbrack BS95\rbrack\,90 & $0.333 \pm 0.048$ & $0.823 \pm 0.046$ & $0.490 \pm 0.045$\\
Kron\,44 & $0.262 \pm 0.014$ & $0.762 \pm 0.013$ & $0.500 \pm 0.013$\\
NGC\,361 & $0.228 \pm 0.016$ & $0.692 \pm 0.015$ & $0.464 \pm 0.011$\\
NGC\,411 & $0.300 \pm 0.024$ & $0.787 \pm 0.024$ & $0.487 \pm 0.022$\\
NGC\,416 & $0.274 \pm 0.031$ & $0.743 \pm 0.028$ & $0.469 \pm 0.028$\\
NGC\,419 & $0.254 \pm 0.025$ & $0.751 \pm 0.025$ & $0.497 \pm 0.025$\\
\lbrack RZ2005\rbrack\,194 & $0.295 \pm 0.030$ & $0.852 \pm 0.037$ & $0.557 \pm 0.031$\\
ESO\,29-46 & $0.203 \pm 0.041$ & $0.711 \pm 0.042$ & $0.508 \pm 0.055$\\
ESO\,29-48 & $0.239 \pm 0.029$ & $0.699 \pm 0.022$ & $0.460 \pm 0.025$\\
NGC\,643 & $0.333 \pm 0.036$ & $0.669 \pm 0.056$ & $0.336 \pm 0.051$\\
Lindsay\,113 & $0.210 \pm 0.013$ & $0.698 \pm 0.013$ & $0.488 \pm 0.012$\\
\hline
\end{tabular}
\end{table}

\begin{table*}
\caption{The derived RC absolute magnitude and  colour of the Milky Way target star clusters.}
\label{tab:RC_am_colors}
\begin{tabular}{lcccccc}
\hline
Cluster name & $M_{J}$ & $M_{H}$ & $M_{K_{S}}$ & $J - H$ & $J - K_{S}$ & $H - K_{S}$\\
\hline
NGC\,188 & $-1.127 \pm 0.017$ & $-1.698 \pm 0.016$ & $-1.805 \pm 0.023$ & $0.571 \pm 0.020$ & $0.679 \pm 0.026$ & $0.108 \pm 0.026$\\
NGC\,752$^a$ & $-0.809 \pm 0.080$ & $-1.269 \pm 0.082$ & $-1.385 \pm 0.081$ & $0.460 \pm 0.115$ & $0.576 \pm 0.114$ & $0.116 \pm 0.115$\\
NGC\,1817 & $-1.201 \pm 0.059$ & $-1.595 \pm 0.063$ & $-1.684 \pm 0.065$ & $0.394 \pm 0.084$ & $0.483 \pm 0.086$ & $0.089 \pm 0.089$\\
M\,37$^a$ & $-1.969 \pm 0.060$ & $-2.311 \pm 0.104$ & $-2.468 \pm 0.063$ & $0.342 \pm 0.119$ & $0.499 \pm 0.086$ & $0.157 \pm 0.121$\\
NGC\,2204 & $-1.404 \pm 0.045$ & $-1.847 \pm 0.034$ & $-1.957 \pm 0.034$ & $0.443 \pm 0.036$ & $0.554 \pm 0.035$ & $0.111 \pm 0.020$\\
NGC\,2243 & $-1.506 \pm 0.044$ & $-1.936 \pm 0.045$ & $-1.993 \pm 0.040$ & $0.431 \pm 0.045$ & $0.487 \pm 0.040$ & $0.057 \pm 0.042$\\
Haffner\,2 & $-2.028 \pm 0.166$ & $-2.440 \pm 0.167$ & $-2.549 \pm 0.167$ & $0.411 \pm 0.028$ & $0.520 \pm 0.035$ & $0.109 \pm 0.032$\\
NGC\,2360 & $-1.244 \pm 0.058$ & $-1.564 \pm 0.059$ & $-1.607 \pm 0.066$ & $0.320 \pm 0.082$ & $0.363 \pm 0.087$ & $0.043 \pm 0.088$\\
Melotte\,66 & $-1.326 \pm 0.043$ & $-1.834 \pm 0.038$ & $-1.924 \pm 0.038$ & $0.508 \pm 0.026$ & $0.598 \pm 0.028$ & $0.090 \pm 0.019$\\
Berkeley\,39 & $-1.275 \pm 0.060$ & $-1.801 \pm 0.058$ & $-1.976 \pm 0.057$ & $0.526 \pm 0.032$ & $0.701 \pm 0.031$ & $0.175 \pm 0.027$\\
NGC\,2477 & $-0.943 \pm 0.027$ & $-1.392 \pm 0.024$ & $-1.494 \pm 0.027$ & $0.449 \pm 0.036$ & $0.551 \pm 0.038$ & $0.102 \pm 0.036$\\
NGC\,2506 & $-1.222 \pm 0.021$ & $-1.721 \pm 0.022$ & $-1.790 \pm 0.019$ & $0.499 \pm 0.021$ & $0.568 \pm 0.019$ & $0.069 \pm 0.019$\\
NGC\,2527$^a$ & $-1.215 \pm 0.070$ & $-1.619 \pm 0.072$ & $-1.767 \pm 0.074$ & $0.404 \pm 0.100$ & $0.552 \pm 0.102$ & $0.148 \pm 0.103$\\
NGC\,2682$^a$ & $-1.119 \pm 0.016$ & $-1.607 \pm 0.010$ & $-1.738 \pm 0.010$ & $0.488 \pm 0.018$ & $0.619 \pm 0.018$ & $0.131 \pm 0.013$\\
NGC\,3680$^a$ & $-1.348 \pm 0.097$ & $-1.855 \pm 0.090$ & $-1.980 \pm 0.090$ & $0.507 \pm 0.132$ & $0.631 \pm 0.132$ & $0.124 \pm 0.126$\\
NGC\,3960$^a$ & $-1.050 \pm 0.097$ & $-1.562 \pm 0.098$ & $-1.647 \pm 0.100$ & $0.512 \pm 0.136$ & $0.597 \pm 0.138$ & $0.085 \pm 0.139$\\
NGC\,5822 & $-1.050 \pm 0.097$ & $-1.562 \pm 0.098$ & $-1.647 \pm 0.100$ & $0.512 \pm 0.136$ & $0.597 \pm 0.138$ & $0.085 \pm 0.139$\\
NGC\,6134 & $-0.981 \pm 0.044$ & $-1.331\pm 0.024$ & $-1.498 \pm 0.044$ & $0.350 \pm 0.049$ & $0.517 \pm 0.061$ & $0.167 \pm 0.049$\\
IC\,4651 & $-1.098 \pm 0.048$ & $-1.682\pm 0.062$ & $-1.794 \pm 0.059$ & $0.584 \pm 0.078$ & $0.696 \pm 0.076$ & $0.112 \pm 0.085$\\
NGC\,6791 & $-1.408 \pm 0.013$ & $-1.948 \pm 0.026$ & $-2.095 \pm 0.025$ & $0.540 \pm 0.020$ & $0.687 \pm 0.017$ & $0.148 \pm 0.017$\\
NGC\,6819 & $-1.486 \pm 0.013$ & $-1.952 \pm 0.019$ & $-2.049 \pm 0.015$ & $0.466 \pm 0.021$ & $0.563 \pm 0.018$ & $0.097 \pm 0.022$\\
NGC\,7789 & $-1.145 \pm 0.020$ & $-1.639 \pm 0.020$ & $-1.750 \pm 0.021$ & $0.493 \pm 0.028$ & $0.605 \pm 0.028$ & $0.111 \pm 0.028$\\
\hline
\multicolumn{7}{l}{$^a$ Clusters that simply adopt the average of the magnitude of RC candidate stars}
\end{tabular}
\end{table*}

\subsection{The Results of Multiple Regression Analysis}

\begin{figure*}
    \includegraphics{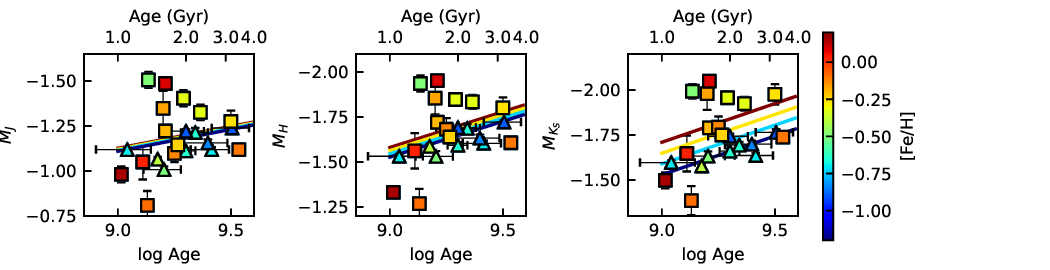}
    \caption{The same figure as Fig.~\ref{fig:AM_age} but with the best-fitting relations for four metallicities (from blue to brown, $-1.0$, $-0.67$, $-0.33$, $0.0$~dex).}
    \label{fig:AM_age_with_fit}
\end{figure*}

\begin{figure*}
    \includegraphics{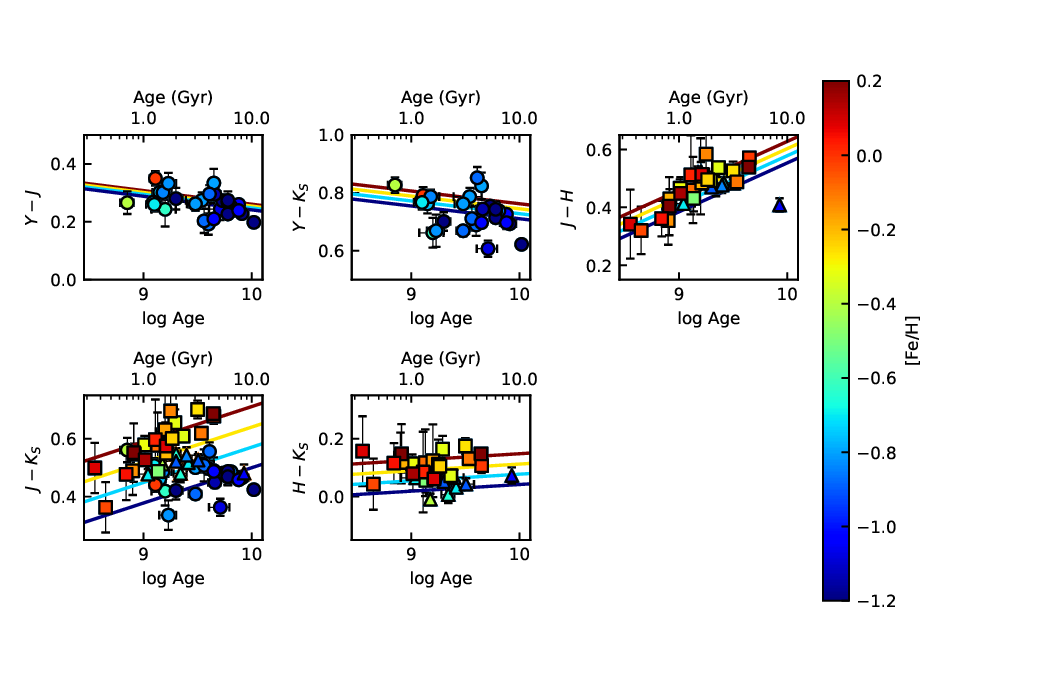}
    \caption{The same figure as Fig.~\ref{fig:color_age} but with the best-fitting relations for four metallicities (from blue to brown, $-1.20$, $-0.73$, $-0.27$, $0.20$~dex).}
    \label{fig:color_age_with_fit}
\end{figure*}

The following values were derived as the best fit results of absolute magnitude,
\begin{align}
M_{J} = (&-0.243 \pm 0.200) (\log(t) - 9.300) \notag \\
&- (0.012 \pm 0.102) ([\mathrm{Fe}/\mathrm{H}] + 0.50) - (1.188 \pm 0.038) \label{eq:M_J}\\
M_{H} = (&-0.397 \pm 0.198) (\log(t) - 9.300) \notag \\ &- (0.038 \pm 0.101) ([\mathrm{Fe}/\mathrm{H}] + 0.50) - (1.671 \pm 0.038) \label{eq:M_H}\\
M_{K_{S}} = (&-0.430 \pm 0.197) (\log(t) - 9.300) \notag \\ &- (0.129 \pm 0.100) ([\mathrm{Fe}/\mathrm{H}] + 0.50) - (1.745 \pm 0.038) \label{eq:M_K},
\end{align}
and colour,
\begin{align}
Y - J = (&-0.047 \pm 0.033) (\log(t) - 9.300)\notag \\
&+ (0.014 \pm 0.031) ([\mathrm{Fe}/\mathrm{H}] + 0.50) + (0.281 \pm 0.012) \label{eq:Y_J}\\
Y - K_{S} = (&-0.044 \pm 0.050) (\log(t) - 9.300)\notag \\ &+ (0.037 \pm 0.048) ([\mathrm{Fe}/\mathrm{H}] + 0.50) + (0.759 \pm 0.019) \label{eq:Y_K}\\
J - H = &(0.169 \pm 0.033) (\log(t) - 9.300)\notag \\
&+ (0.053 \pm 0.027) ([\mathrm{Fe}/\mathrm{H}] + 0.50) + (0.472 \pm 0.010) \label{eq:J_H}\\
J - K_{S} = &(0.122 \pm 0.032) (\log(t) - 9.300)\notag \\
&+ (0.151 \pm 0.023) ([\mathrm{Fe}/\mathrm{H}] + 0.50) + (0.525 \pm 0.008) \label{eq:J_K}\\
H - K_{S} = &(0.023 \pm 0.027) (\log(t) - 9.300)\notag \\
&+ (0.076 \pm 0.022) ([\mathrm{Fe}/\mathrm{H}] + 0.50) + (0.078 \pm 0.008) \label{eq:H_K}.
\end{align}
The equations derived for the absolute magnitude are compared for the four metallicities in Fig.~\ref{fig:AM_age_with_fit} and for the colour in Fig.~\ref{fig:color_age_with_fit}. We calculated the coefficients of determination to assess the goodness of fit. The coefficient of determination, adjusted for degrees of freedom (adjusted $R^{2}$), is given by
\begin{equation}
\mathrm{adjusted}\ R^{2} = 1 - \frac{\sum_{i}((M_{\lambda} - m_{\lambda, i}) - f_{i})^2 / (N - p - 1)}{\sum_{i}(M_{\lambda, i} - \overline{M_{\lambda}})^2 / (N  - 1)},
\end{equation}
where $M_{\lambda}$ is the RC colour of observational data, $f$ is the colour from equations~(\ref{eq:M_J})--(\ref{eq:M_K}), $\overline{M_{\lambda}}$ is the average of $M_{\lambda}$, $N$ is the number of sample star clusters, and $p$ is the number of explanatory variables (three in this time). The calculated values of adjusted $R^{2}$ are 0.066, 0.165, and 0.234 for $M_{J}$, $M_{H}$, and $M_{K_{S}}$, respectively. These values are close to zero, indicating that the fitting results differ little from simply taking the average. The adjusted $R^{2}$ values for colour are 0.204, 0.165, 0.477, 0.433, and 0.306 for $Y - J$, $Y - K_{S}$, $J - H$, $J - K_{S}$, and $H - K_{S}$, respectively.

\subsection{Population effects}
\label{sec:PE}
Multiple regression analysis indicates that the older RC stars are brighter in the 1--4\,Gyr range for all $M_{J}$, $M_{H}$, and $M_{K_{S}}$, although the adjusted $R^{2}$ values are lower. This result is consistent with the LMC-only results of \citet{OIN2019} and the theoretical predictions of \citet{G2016}. The rapid brightening of the young RC with an age of 0.35 Gyr, although only in one cluster, is also in agreement with theoretical predictions.

Regarding the effect of metallicity on the NIR absolute magnitude of RC stars, no clear dependence has been confirmed. The best-fit results from the multiple regression analysis indicate brighter for higher metallicity within the 1--4 Gyr range for all $JHK_{S}$ bands, although the dependence is not significant when accounting for uncertainties. The weak metallicity dependence of the NIR absolute magnitude is consistent with observational results \citep{A2000,G2008,LJP2012} and theoretical predictions by \citet{G2016}.

In terms of colour, $J - H$ and $J - K_{S}$ exhibit age dependence, with older RC stars appearing redder, whereas $Y - J$, $Y - K_{S}$, and $H - K_{S}$ do not show a strong dependence. Conversely, metallicity dependence indicates redder colour at higher metallicity for all colour, consistent with theoretical predictions, although the trend is weaker for $Y - J$ and $Y - K_{S}$. However, $Y - J$ and $Y - K_{S}$ data are available only for the SMC, and the trend may not be observed due to the limited age and metallicity range covered by these data. As shown in Fig.~\ref{fig:Age_and_Metallicity}, the age-metallicity relationship is evident only the sample of the SMC clusters, where older star clusters have lower metallicity and younger star clusters have higher metallicity. The theoretical models predict that RC stars in the age range (0.5 $\lesssim$ age (Gyr) $\lesssim$ 8) become redder as they get older or have higher metallicity. In contrast, older RC stars are predicted to become bluer for young metal-rich RC stars (age $\lesssim$ 0.5\, Gyr, [Fe/H] $\gtrsim$ 0.05) and old metal-poor RC stars (age $\gtrsim$ 8\, Gyr, [Fe/H] $\lesssim$ -0.40). If samples from a single galaxy following the age-metallicity relation closely, the population effects are likely to cancel each other out. For $J - H$, the strong age dependence may result from the lack of SMC samples covering low-metallicity clusters.

\subsection{Comparison with theoretical models} 
We compare the observed results not only with the model of \citet{G2016}, but also with other theoretical models. We used three isochrone models for comparison: a Bag of Stellar Tracks and Isochrones \citep[BaSTI, ][]{PCS2004}, PARSEC version~1.2S evolutionary tracks \citep{BMG2012} and canonical two-part-power law IMF of \citet{K2001,K2002}, and MESA Isochrones \& Stellar Tracks (MIST) version~1.2 \citep{PBD2011,PCM2013,PMS2015,PSB2018,D2016,CDC2016}. Although other isochrone models are available, we used these models because they can perform calculations up to and beyond the helium burning stage, and they enable plotting isochrones across a wide range of age and metallicity. To define the mass range of RC stars in the isochrone models, we identify the region where the absolute magnitude remains nearly constant during the transition from the red giant branch (RGB) to the asymptotic giant branch. The lower mass boundaries of the RC stars are set at the minimum initial mass above the tip of the RGB for which the derivative of luminosity with respect to mass ($dL/dM$) becomes positive, indicating the beginning of a relatively flat luminosity region. The upper boundaries are defined as the lowest mass at which $dL/dM$ exceeds the mean value plus three times the standard deviation of $dL/dM$, evaluated over the interval from the lower boundary to a higher reference mass. This criterion provides a consistent and objective way to isolate the RC region from neighbouring evolutionary phases. The median absolute magnitude of stars within this mass interval is adopted as the representative RC magnitude for each age and metallicity in the models. Specifically, let $M_{1}$ be the minimum mass at which $dL/dM > 0$ above the tip of the RGB, and $M_{2}$ the minimum mass satisfying $dL/dM > \langle dL/dM\rangle + 3\sigma$ within $[M_{1}, M_{\mathrm{ref}}]$; the RC magnitude is defined as the median absolute magnitude over [$M_{1}$, $M_{2}$].

\begin{figure*}
    \includegraphics{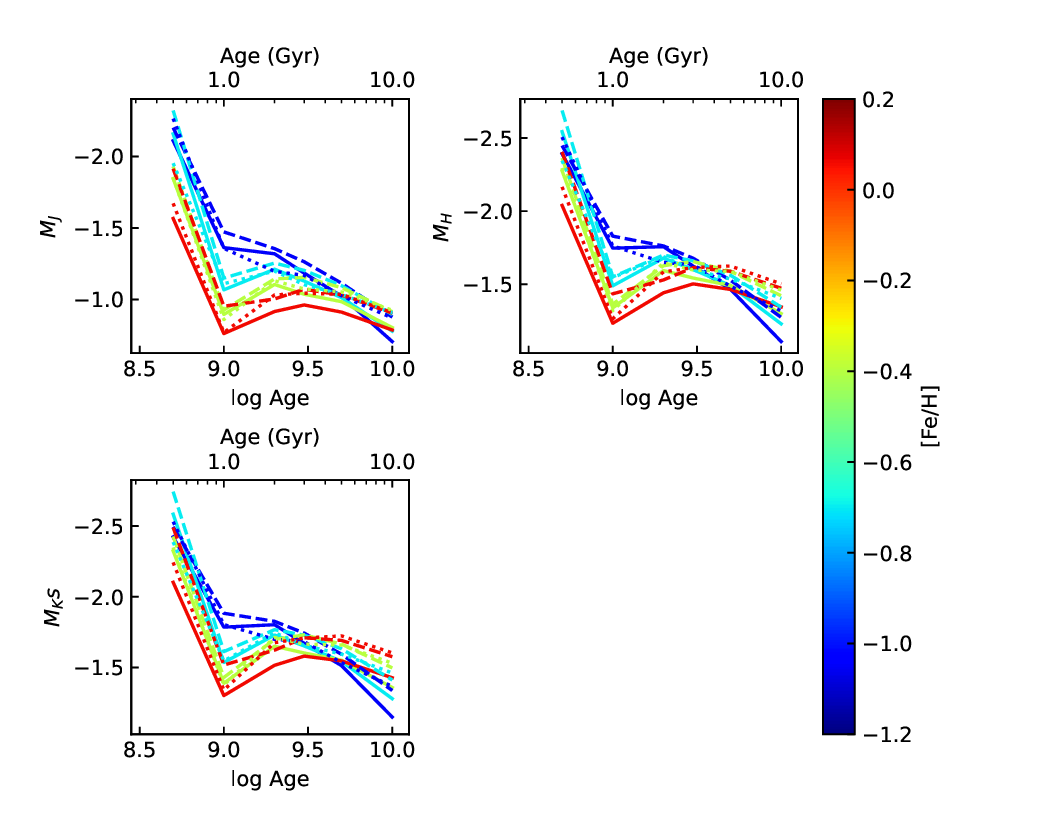}
    \caption{Comparison of the three isochrone models for four metallicities (from blue to brown, $-1.05$, $-0.70$, $-0.40$, $0.06$~dex). Solid lines are BaSTI, dashed lines are PARSEC, and dotted lines are MIST.}
    \label{fig:AM_Model_comparison}
\end{figure*}

\begin{figure*}
    \includegraphics{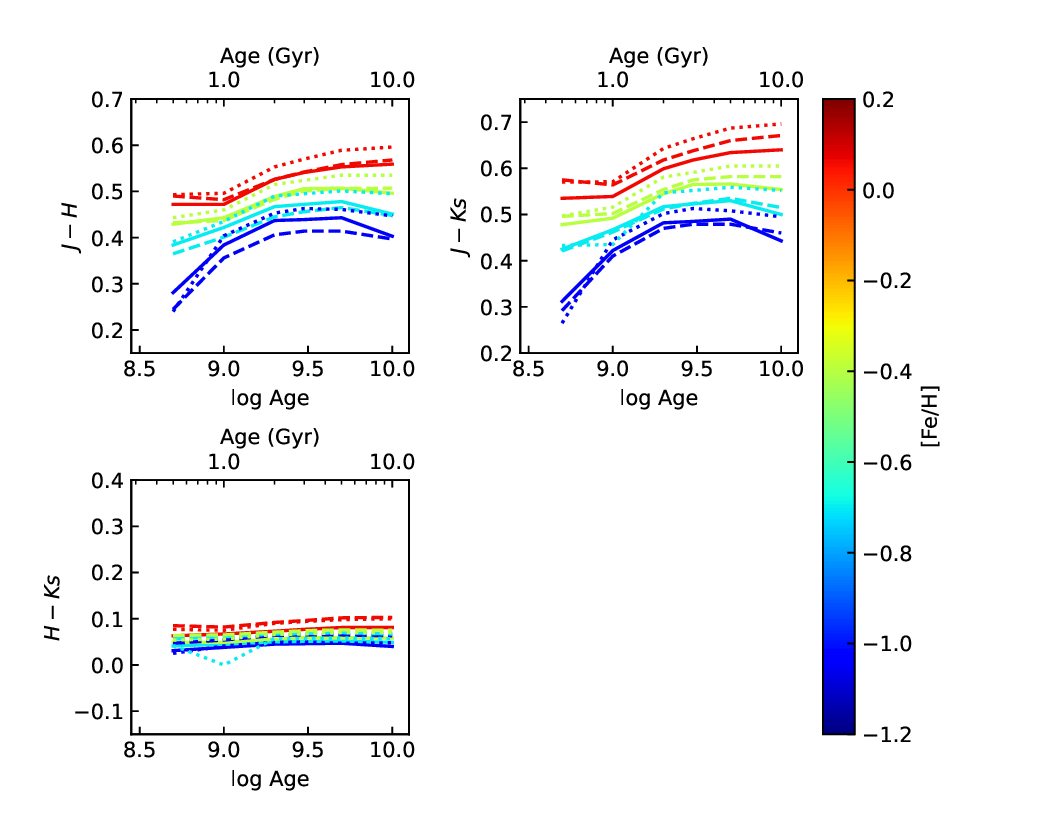}
    \caption{Comparison of the three isochrone models for four metallicities (from blue to brown, $-1.05$, $-0.70$, $-0.40$, $0.06$~dex). Solid lines are BaSTI, dashed lines are PARSEC, and dotted lines are MIST. Triangles are BaSTI, Circles are CMD 3.7, and Squares are MIST.}
    \label{fig:Color_model_comparison}
\end{figure*}

No significant differences are observed in the trend of the population effect on absolute magnitude as shown in Fig.~\ref{fig:AM_Model_comparison}. When comparing absolute values, PARSEC appears to be slightly brighter. The population effect on RC colour shows little differences in trends between models. RC stars are commonly redder if they are older or more metal rich as depicted in Fig.~\ref{fig:Color_model_comparison}. Additionally, the population effect is negligible for $H - K_{S}$. However, some differences are observed in the absolute values.

\begin{figure*}
    \includegraphics{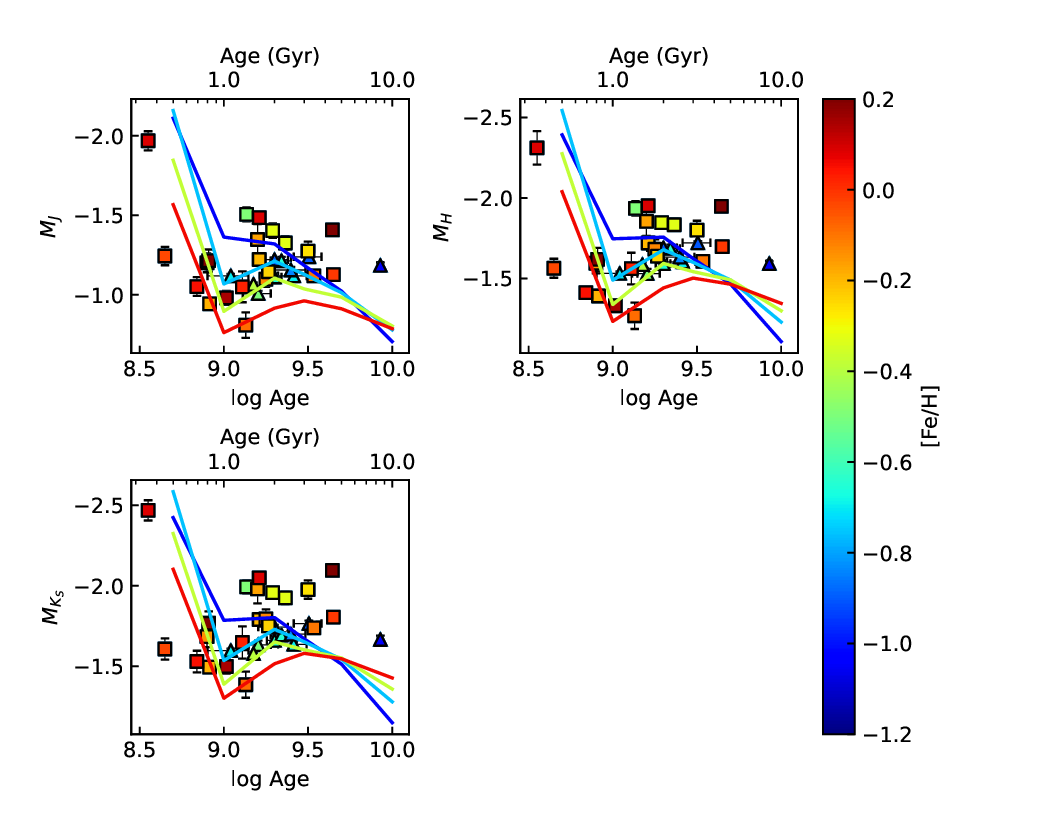}
    \caption{Comparison of absolute magnitude between the observed results and the BaSTI isochrone models for four metallicities (from blue to brown, $-1.05$, $-0.70$, $-0.40$, $0.06$~dex).}
    \label{fig:Obs_and_BaSTI_AM}
\end{figure*}

\begin{figure*}
    \includegraphics{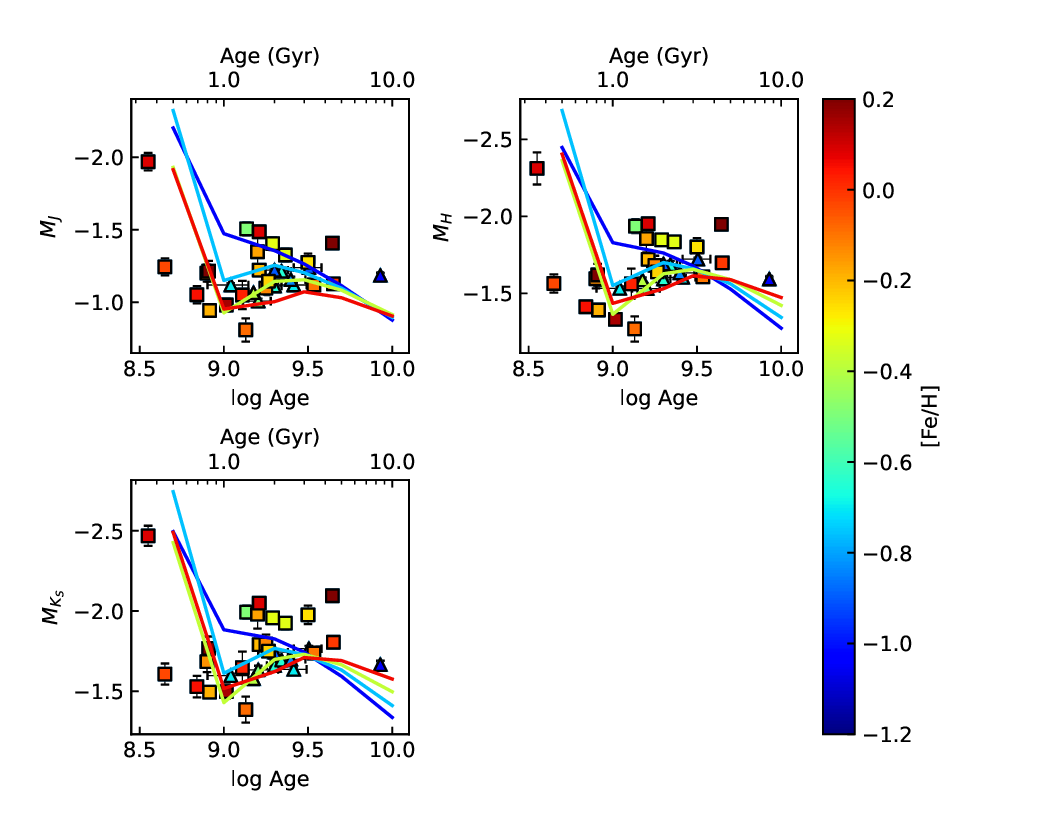}
    \caption{Comparison of absolute magnitude between the observed results and the PARSEC isochrone models for four metallicities (from blue to brown, $-1.05$, $-0.70$, $-0.40$, $0.06$~dex).}
    \label{fig:Obs_and_CMD37_AM}
\end{figure*}

\begin{figure*}
    \includegraphics{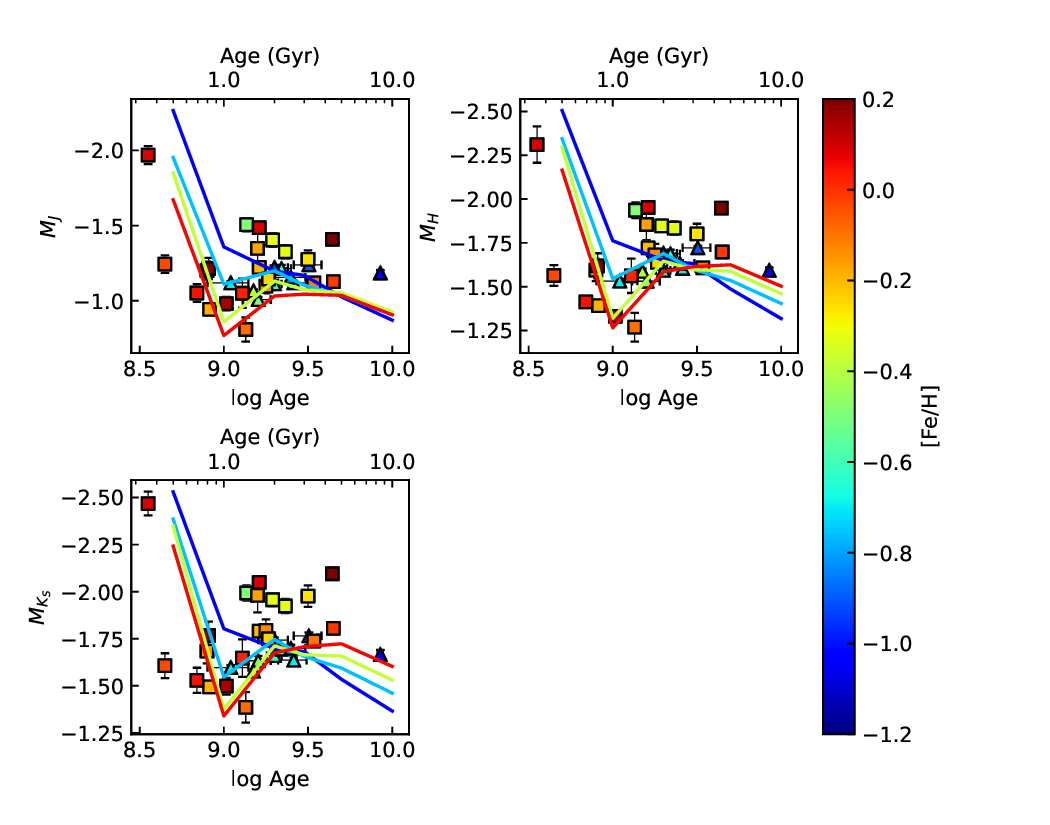}
    \caption{Comparison of absolute magnitude between the observed results and the MIST isochrone models for four metallicities (from blue to brown, $-1.05$, $-0.70$, $-0.40$, $0.06$~dex).}
    \label{fig:Obs_and_MIST_AM}
\end{figure*}

\begin{figure*}
    \includegraphics{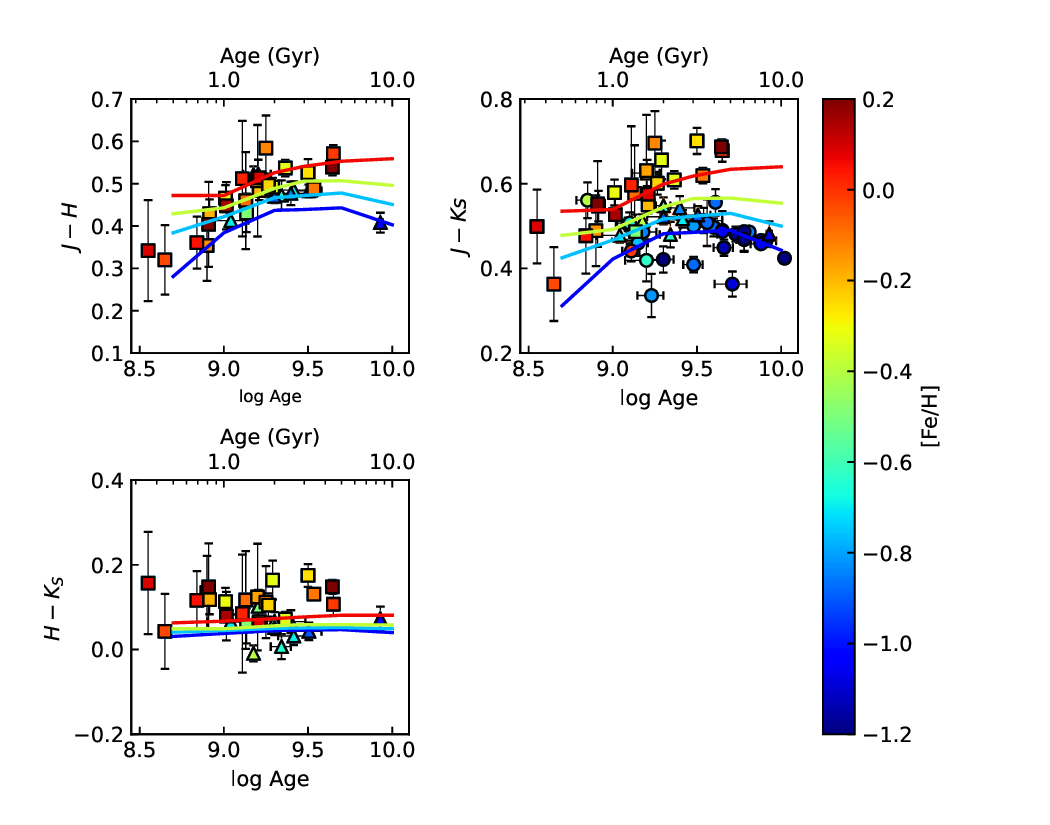}
    \caption{Comparison of colour between the observed results and the BaSTI isochrone models for four metallicities (from blue to brown, $-1.05$, $-0.70$, $-0.40$, $0.06$~dex).}
    \label{fig:Obs_and_BaSTI_color}
\end{figure*}

\begin{figure*}
    \includegraphics{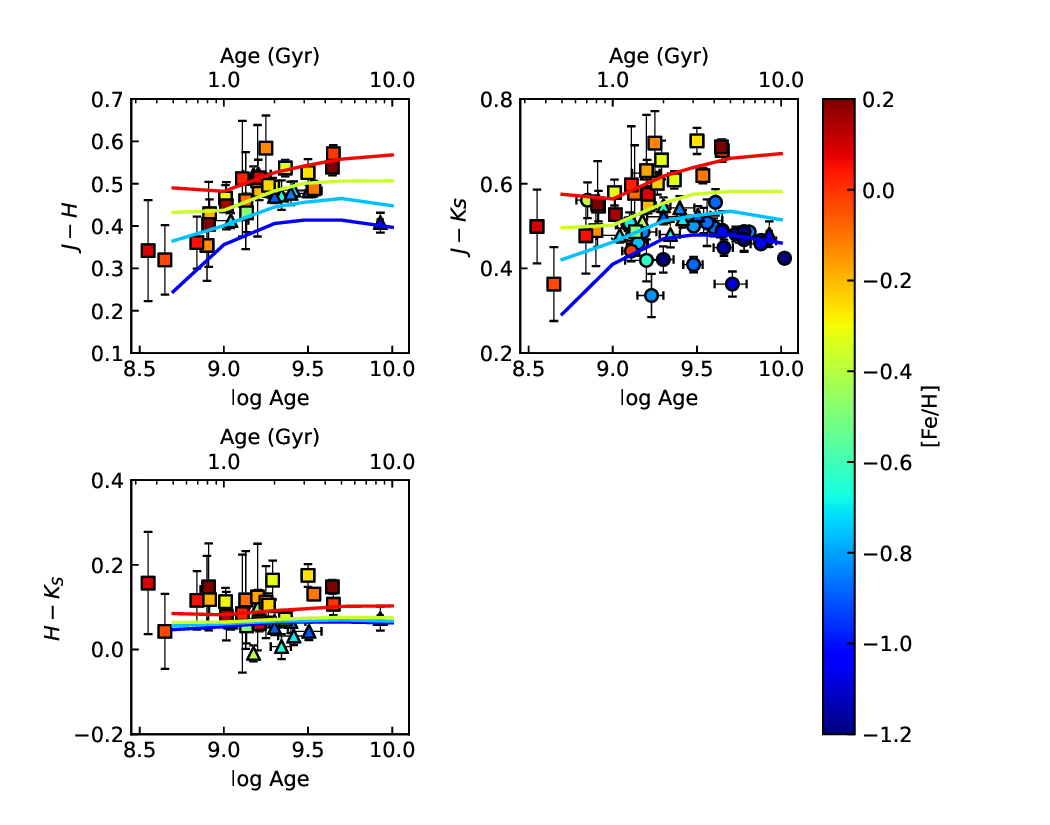}
    \caption{Comparison of colour between the observed results and the PARSEC isochrone models for four metallicities (from blue to brown, $-1.05$, $-0.70$, $-0.40$, $0.06$~dex).}
    \label{fig:Obs_and_CMD37_color}
\end{figure*}

\begin{figure*}
    \includegraphics{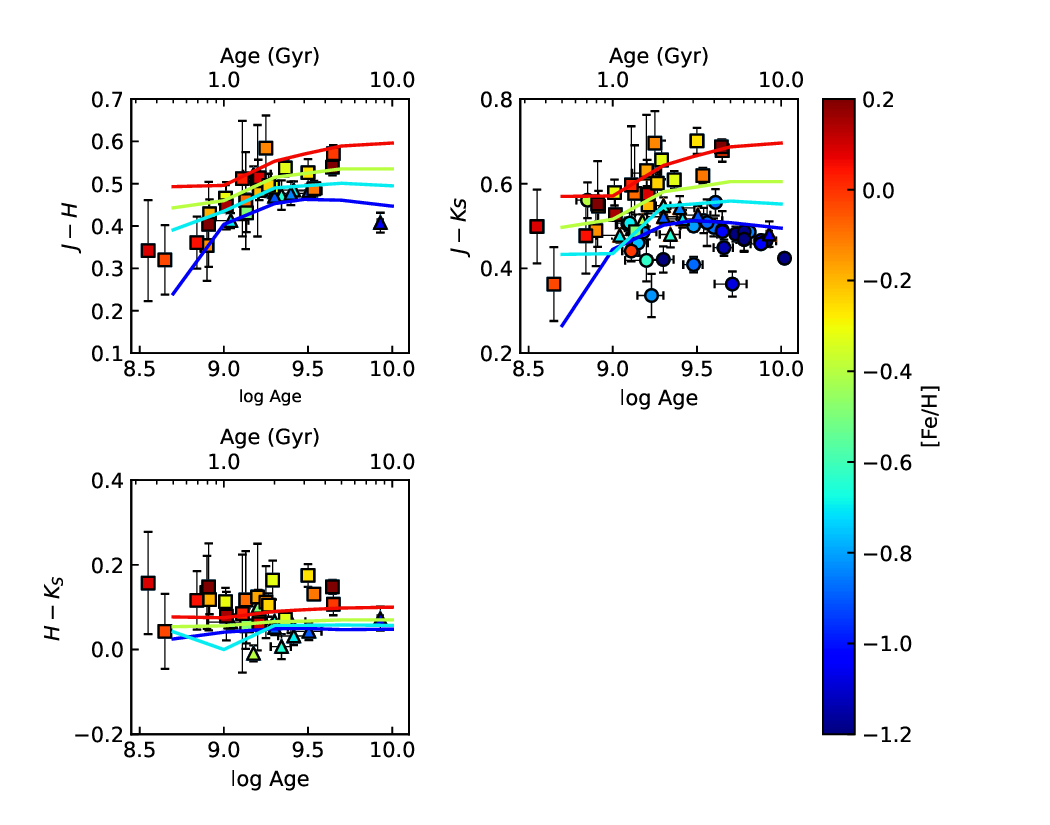}
    \caption{Comparison of absolute magnitude between the observed results and the MIST isochrone models for four metallicities (from blue to brown, $-1.05$, $-0.70$, $-0.40$, $0.06$~dex).}
    \label{fig:Obs_and_MIST_color}
\end{figure*}

Figs.~\ref{fig:Obs_and_BaSTI_AM} through \ref{fig:Obs_and_MIST_AM} compare the three models with our results for absolute magnitude, while Figs.~\ref{fig:Obs_and_BaSTI_color} through \ref{fig:Obs_and_MIST_color} compare them for colour. The comparison results indicate that the age dependence is well reproduced. For colour, both the trends and absolute values show good agreement with observations and the models. In PARSEC, the model for metal-poor RC stars appears slightly bluer in $J - H$. The $H - K_{S}$ values deviate slightly due to the weak dependence, but they remain consistent within the range of observational uncertainties.

\subsection{Comparison with observational results in the solar neighbourhood}

The mean absolute magnitude of RC stars in the solar neighbourhood is $M_{J} \sim -1.0$, $M_{H} \sim -1.5$, and $M_{K_{S}} \sim -1.6$ \citep{A2000, G2008, LJP2012, RBA2018, POB2020}, although minor variations exist between studies. The simple averages of our results for target star clusters are $M_{J} = -1.190 \pm 0.034$, $M_{H} = -1.678 \pm 0.035$, and $M_{K_{S}} = -1.766 \pm 0.037$, which are systematically brighter by 0.15--0.20 compared to the solar neighbourhood. One possible explanation for this discrepancy is the overestimation of distances derived from Gaia's parallaxes for relatively distant star clusters in the Milky Way. For instance, \citet{POB2020} showed that the absolute magnitude of RC stars, determined by directly using the inverse of parallaxes as the distances, is about 0.10--0.15 mag brighter than that derived by adopting the results of \citet{BRF2018}, which used the galactic coordinates as a prior, considering the Galactic model. However, star clusters with brighter RC stars (NGC\, 2204, NGC\, 2243, NGC\, 6791, and NGC\, 6819) exhibit distances comparable to those obtained via isochrone fitting as employed by \citet{vHG2007}, suggesting that the difference in absolute magnitude may arise from the population differences between the star clusters and the solar neighbourhood. As shown by the results of our results and as predicted by theory, the population effect on absolute magnitude is greater than the effect of age difference. Although the solar neighbourhood differs from star clusters in that it contains stars of various ages, it is possible to estimate the typical age from the absolute magnitude by assuming the metallicity. Applying the formula derived in this study to the absolute magnitude of the solar neighbourhood with [Fe/H] = 0, we obtain $\log(t) \sim 8.5$ in the $J$-band and $\log(t) \sim 8.8$ in the $H$- and $K_{S}$-bands. These values lie near the edge of the age range of the star clusters used for the fitting in this study suggesting that the local RC stars mainly originate from either younger or older populations. Expanding the age range of sample RC stars would be greatly facilitated if photometric data for extragalactic star clusters become available in the future. In addition, asteroseismology of the local RC stars is a promising approach to investigate their NIR brightness.

The average colour of RC stars in the solar neighbourhood is $J - H \sim 0.5$, $J - K_{S} \sim 0.6$, and $H - K_{s} \sim 0.1$, respectively. The simple averages of our results are $J - H = 0.469 \pm 0.011$, $J - K_{S} = 0.517 \pm 0.010$, and $H - K_{S} = 0.092 \pm 0.008$, which are bluer than the corresponding values for the solar neighbourhood for $J - H$ and $J - K_{S}$. When the samples are divided into RC stars from the Milky Way star clusters, the LMC star clusters and the SMC star clusters, $J - K_{S}$ is $0.577 \pm 0.018$ in the Milky Way, $0.523 \pm 0.014$ in the LMC and $0.468 \pm 0.009$ in the SMC, respectively. This reflects population effects, especially for the metallicity dependence. The mean colour of the Milky Way star clusters, which have metallicity close to that of the Sun, is similar to the mean colour of the RC stars in the solar neighbourhood, while the mean colour of the SMC star clusters, which are dominated by low metallicity, is clearly bluer than those in the solar neighbourhood. This colour difference is consistent with the population effect trend identified in this study. On the other hand, the $H - K_{S}$ values are almost identical for both our samples and RC stars in the solar neighbourhood, which is also consistent with our results.

\section{Conclusions}
\label{sec:conclusions}
Using star clusters from the SMC and the Milky Way Galaxy, we investigate age and metallicity dependence of the absolute magnitude and colour of RC stars in the near-infrared within a previously unexplored parameter range. Multiple regression analysis was performed to examine the population effect on absolute magnitude. The results are consistent with the theoretical models, indicating that for RC stars in the 1-4 Gyr range, there is an age dependence with older RC stars being brighter, while the metallicity dependence is weak. For more accurate validation, particularly regarding the metallicity dependence, samples with precisely determined age, metallicity, and distance are required. Regarding colour, the trends for both age and metallicity align with those predicted by theoretical models. In particular, for $J - K_{S}$, the complete set of the LMC, the SMC and the Milky Way star cluster allowed us to extensively cover the parameter space, thus confirming both the age and metallicity dependence.

\section*{Acknowledgements}

Based on data products created from observations collected at the European Organisation for Astronomical Research in the Southern Hemisphere under ESO programme 179.B-2003. This publication makes use of data products from the Two Micron All Sky Survey, which is a joint project of the University of Massachusetts and the Infrared Processing and Analysis Center/California Institute of Technology, funded by the National Aeronautics and Space Administration and the National Science Foundation. This work was supported by JSPS KAKENHI Grant Number 23K20860.

\section*{Data Availability}

The PSF photometry data of the VMC survey is available from ESO Catalogue Facility (https://www.eso.org/qi/catalogQuery/index/340). The 2MASS photometry is available from NASA/IPAC Infrared Science Archive (https://irsa.ipac.caltech.edu/Missions/2mass.html).



\bibliographystyle{mnras}
\bibliography{reference} 








\bsp	
\label{lastpage}
\end{document}